\documentclass[a4paper, 12pt]{article}

\usepackage{array,bbm}
\usepackage{booktabs}
\usepackage{rotating}
\usepackage{amsmath}
\usepackage{amsfonts}
\usepackage{color}
\usepackage{graphics}

\usepackage{undertilde}
\usepackage{titlesec,blindtext}
\usepackage{enumitem}
\usepackage{placeins}
\usepackage{natbib}
\usepackage{listings}
\usepackage{datetime}
\usepackage{multirow}
\usepackage{hyperref}
\usepackage{verbatim}

\usepackage{xr}
\def\rev#1{{\color{red} #1}}

\begin{document}
	\begingroup
\centering
\large \textbf{Conditional analysis for mixed covariates, with application to feed intake of lactating sows}\\[2em]
S.Y. Park$\null^{1,*}$, C. Li$\null^{2,**}$, S.M. Mendoza Benavides$\null^{3}$, E. van Heugten$\null^{3}$ and A.M. Staicu$\null^{2}$ \\[1em]

{\small\it $\null^{1}$Eli Lilly and Company, Indianapolis, IN 46285, USA}\\
{\small\it $\null^{2}$Department of Statistics, North Carolina State University, Raleigh, NC 27695, USA}\\
{\small\it $\null^{3}$Department of Animal Science, North Carolina State University, Raleigh, NC 27695, USA}\\
{\small\it $\null^{*}$Email: parksoyoung3859@gmail.com} \\
{\small\it $\null^{**}$Email: cli9@ncsu.edu} \\[1em]
{\small \today} \\[1em]
\endgroup

%  pub the summary here
\begin{abstract}
		We propose a novel modeling framework to study the effect of covariates of various types on the conditional distribution of the response. The methodology accommodates flexible model structure, allows for joint estimation of the quantiles at all levels, and provides a computationally efficient estimation algorithm. Extensive numerical investigation confirms good performance of the proposed method. The methodology is motivated by and applied to a lactating sow study, where the primary interest is to understand how the dynamic change of minute-by-minute temperature in the farrowing rooms within a day (functional covariate) is associated with low quantiles of feed intake of lactating sows, while accounting for other sow-specific information (vector covariate).
\end{abstract}

%
%  Please place your key words in alphabetical order, separated
%  by semicolons, with the first letter of the first word capitalized,
%  and a period at the end of the list.
%

%\begin{keywords}
%Conditional Distribution, Functional Data Analysis, Feed Intake of Lactating Sows,  Functional Quantile Regression, Penalized Likelihood.
%\end{keywords}

\section{Introduction}
%START
Many modern applications routinely collect data on study participants comprising scalar responses and covariates of various types, vector, function, image, and the main question of interest is to examine how the covariates affect the response. For example, in our motivating experimental study the goal is to analyze how the minute-by-minute daily temperature and humidity of the farrowing rooms, where sows are placed after giving birth for nursing, affect their feed intake during a lactation period. The covariates consist of temperature profile, humidity, and sow age, where the response is a total amount of daily feed intake of sows.
%in our data application the aim is to study how the min-by-min temperature or humidity during a day in the farrowing rooms, that is rooms where piglets are born and nursed by the sow until they are weaned, affect the feed intake of lactating sows in their first 21 lactation days, while accounting for other sow-specific information. The response is feed intake amount at a day, and covariates are the min-by-min temperature for that day, average humidity, age of the respect saw. 	
A popular approach in these cases is to use a nonparametric framework and assume that the mixed covariates solely affect the mean response; see \cite{CardotFerratySarda1999, james2002generalized, ramsay2006functional, ramsay2002applied, ferraty2006nonparametric, ferraty2009additive, goldsmith2012penalized, McLean2014} and others. However for our application, while it is important to study the mean feed intake, animal scientists are often more concerned with the left tail of the feed intake distribution. This is because low feed intake of lactating sows could lead to many serious issues, including decrease in milk production and negative impact on the sows reproductive system; see, for reference, \citet{quiniou1999influence, renaudeau2001effects, st2003economic} among others. 
In this paper we focus on regression models that study the effects of covariates on the entire distribution of response. Our contribution is the development of a modeling framework that accommodates a comprehensive study of various types (vector and functional) of covariates on a scalar response.

Quantile regression models the effect of scalar/vector covariates beyond the mean response, it provides a more comprehensive study of the covariates on the response, and has attracted great interest \citep*{koenker1978regression, koenker2005quantile}. For pre-specified quantile levels, quantile regression models the conditional quantiles of the response as a function of the observed covariates; this approach has been extended more recently to ensure non-crossing of quantile functions \citep{BondellReichWang2010}. Quantile regression has been also extended to handle functional covariates. \cite{cardot2005quantile} discussed quantile regression models by employing a smoothing spline modeling based approach. \cite{kato2012estimation} considered the same problem and used a functional principal component (fPC) based approach. Both papers mainly discussed the case of having a single functional covariate and it is not clear how to extend them to the case where there are multiple functional covariates or mixed covariates (vector and functional). 

More recently, \citet{tang2014partial}, \citet{lu2014functional}, and \citet{yu2015partial} studied quantile regression when the covariates are of mixed types and introduced the partial functional linear quantile regression modeling framework. The first two publications used fPC basis while the last one considered partial quantile regression (PQR) basis. These approaches all are suitable when the interest is studying the effect of covariate at a particular quantile level, and do not handle the study of covariate effects at simultaneous quantile levels due to the well-known crossing-issue.

\cite{ferraty2005conditional} and \cite{chen2012conditional} considered a different perspective and studied the effect of a functional predictor on the quantiles of the response  by modeling the conditional distribution of the response directly. However their approach is limited to one functional predictor. In this paper we fill this gap and propose a unifying modeling framework and estimation technique that allows to study the effect of mixed type covariates (i.e. scalar, vector, and functional) on the conditional distribution of a scalar response in a computationally efficient manner. 

Let $Q_{Y|X(\cdot)}(\tau)$ denote the $\tau$th conditional quantile of $Y$ given a functional covariate $X(\cdot)$, and let $F_{Y|X(\cdot)}(y)$ denote the conditional distribution of $Y$ given $X(\cdot)$. We model the conditional distribution using a generalized function-on-function regression framework, i.e., $F_{Y|X(\cdot)}(y) = E[\mathbbm{1}(Y \leq y) | X(\cdot)] = g^{-1} \{\int X(t) \beta(t, y)dt \}$, where $\mathbbm{1}(\cdot)$ is an indicator function, $g$ is the logit link function, and we study the conditional quantiles by exploiting the relationship between $Q_{Y|X(\cdot)}(\tau)$ and  $F_{Y|X(\cdot)}(y)$ through $Q_{Y|X)(\cdot)}(\tau) = \text{inf}\{y: F_{Y|X(\cdot)}(y) \geq \tau \}$ for $0 < \tau < 1 $. The advantage and contribution of our proposed method  %\citep{park2016conditional} 
mainly come from the following reasons: (1) our modeling approach is spline-based, and as a result it can easily accommodate smooth effects of scalar variables as well as of functional covariates; and (2) our estimation approach is based on a single step function-on-function (or function-on-scalar) penalized regression, which enables efficient implementation by exploiting off-the-shelf software and leads to competitive computations.

The remainder of the paper is structured as follows. Section \ref{sec: method} discusses the details of the proposed method and Section \ref{sec: estimation} describes the estimation procedure and extensions. 
Section \ref{sec: simulation} performs a thorough simulation study evaluating the performance of the proposed method and its competitors. We apply the proposed method to analyze the sow data in Section \ref{sec: sow dat}. We conclude the paper with a discussion in Section \ref{sec: dis}.

\section{Methodology} \label{sec: method}
\subsection{Statistical framework} \label{subsec: modeling framework} 
Let $i$ index subjects, $j$ index repeated measurements, $n$ be the number of subjects, and $m_i$ be the number of observations for subject $i$.
Suppose we observe $\left\{Y_i, \mathbf{X}_{i1}, X_{i2}, \left\{t_{ij}, X_{i3}(t_{ij})\right\}_{1\leq j\leq m_i}\right\}_{1 \leq i \leq n}$, where $Y_i$ is the response, which is a scalar random variable, 
$\mathbf{X}_{i1}$ is a $p$-dimensional vector of nuisance covariates, $X_{i2}$ is a scalar covariate, and $X_{i3}(\cdot)$ is a functional covariate, which %is an independents realization of an underlying stochastic process and 
is assumed to be square-integrable on a closed domain $\mathcal{T}$. 

We propose the following model for the conditional distribution of $Y_i$ given $\mathbf{X}_{i1}$, $X_{i2}$ and $X_{i3}(\cdot)$:
\begin{align} 
	F_{Y_i|\mathbf{X}_{i1}, X_{i2}, X_{i3}(\cdot)}(y) &= E\left\{\mathbbm{1}\left( Y_i \leq y \right) | \mathbf{X}_{i1}, X_{i2}, X_{i3}(\cdot)  \right\}   \label{eq: complex model} \\ 
	&= g^{-1}\big\{\beta_0(y) + \mathbf{X}_{i1}^T\boldsymbol{\beta_1} + X_{i2}\beta_2(y) + \int X_{i3}(t)\beta_3(t,y) dt\big\}, \nonumber
\end{align}
where $F(\cdot)$ denotes the conditional distribution function as before, $g(\cdot)$ is a known, monotone link function, namely the logit link function defined as $g(x) = \log\{x/(1-x)\}$ for arbitrary scalar $x \in [0,1]$, 
$\beta_0(\cdot)$ is an unknown and smooth functional intercept,
$\boldsymbol{\beta_1}$ is a $p$-dimensional
parameter capturing the linear additive effect of the covariate vector $\mathbf{X}_{i1}$,
$\beta_2(\cdot)$ is an unknown and smooth function, and $\beta_3(\cdot,\cdot)$ is an unknown and smooth bivariate function. Here, the effect of the nuisance covariates $\mathbf{X}_{i1}$ is $\boldsymbol{\beta_{1}}$, it is assumed to be constant over $y$ while the smooth intercept $\beta_0(y)$ is $y$-variant. The effect of $X_{i2}$ is $\beta_2(y)$, which varies smoothly over $y$;
$\beta_3(\cdot, y)$ quantifies $y$-variant linear effects of the covariate $X_{i3}(\cdot)$. 
If the parameter function $\beta_2(\cdot)$ is zero then the covariate $X_{i2}$ has no effect on the distribution of the response $Y_i$, which is equivalent to $X_{i2}$ having no effect on any quantile level of $Y_{i}$. Similarly, it is easy to see that a null effect, say $\beta_3(\cdot, \cdot) \equiv 0$, is equivalent to the case that the functional covariate $X_{i3}(\cdot)$ has no effect on any quantile level of the response.
\cite{chen2012conditional} (CM, henceforth) considered a similar model, however, their approach is restrictive to a single functional covariate. We discuss the differences between their method and ours in Section \ref{subsec: basis expansion}
%Their estimation is based on functional principal component bases. %and a direct extension to accommodate mixed covariates is computationally expensive.

To explain our ideas, we consider the case that the functional covariates are observed without noise %noiseless and observed 
on a fine, regular and common grid of sampling points, i.e., $t_{ij} = t_j$ with $j=1,\cdots,m$ for all $i$. Bear in mind, this assumption is made for illustration only, and our framework can be extended to more general cases, 
including settings where the functional covariate is observed with noise and at irregular sampling points,
see Section \ref{subsec: Extension to sparse and noisy functional covariates}.

%	First, we focus on the case of \rev{having} a scalar covariate $X$ \rev{only; the data structure that we consider in this subsection is the same as one in typical quantile regression. Specifically,}
%\rev{ From Cai: (consider moving this part from real data section as I find it difficult to explain parity which is too specific, therefore, I'm thinking probably we also need to move some description of data from that section to introduction part and delete the corresponding part. Or we simply don't need to bother.) In our motivating problem, exploratory analysis of the feed intake behavior of the sows suggest similarities for the sows with parity greater than one (who are at their third pregnancy or higher); thus we use a parity indicator instead of the actual parity of the sow. The parity indicator $P_i$ is defined as one, if the ith sow has parity one and zero otherwise. As emphasized throughout the paper, the existing literature on quantile regression is not suitable to incorporate covariates of different types, as it is the case here.}
%Specifically let the data be $\{(X_i, Y_i):i=1,\ldots, n\}$, where $X_i$ and $Y_i$ are independent realizations of real-valued scalar random variables $X$ and $Y$, respectively. 
%For instance in the sow data application $X$ is the average daily humidity, and $Y$ is daily feed intake.  

\subsection{Modeling of the covariate effects} \label{subsec: basis expansion}
We model $\beta_0(y)$ and $\beta_2(y)$ by using pre-specified, truncated univariate basis. Let $\{B_{0,d_0}(\cdot): d_0=1,\ldots,\kappa_0\}$ and $\{B_{1,d_1}(\cdot): d_1=1,\ldots,\kappa_1\}$ be two bases of dimensions $\kappa_0$ and $\kappa_1$ respectively.
$\beta_0(y) \approx \sum_{d_0=1}^{\kappa_0} B_{0,d_0}(y) \theta_{0,d_0}$ and $\beta_2(y) \approx \sum_{d_1=1}^{\kappa_1} B_{1,d_1}(y) \theta_{1,d_1}$, where $\theta_{0,d_0}$'s and $\theta_{1,d_1}$'s  are unknown basis coefficients. 
We represent $\beta_{3}(t,y)$ using the tensor product of two univariate bases functions, 
$\{B^{t}_{2,d^{}_{2}}(t): d^{}_{2} = 1, \ldots, \kappa_{2,t} \}$ and $\{B^{y}_{2,d^{\prime}_{2}}(y): d^{\prime}_{2} = 1, \ldots, \kappa_{2,y}\}$, where $\kappa_{2,t}$ and $\kappa_{2,y}$ are the bases dimensions; 
$\beta_{3}(t,y) \approx \sum_{d^{}_{2}=1}^{\kappa_{2,t}} \sum_{d^{\prime}_{2}=1}^{\kappa_{2,y}} B^{t}_{2,d^{}_{2}}(t)B^{y}_{2,d^{\prime}_{2}}(y) \theta_{2,d^{}_{2}, d^{\prime}_{2}}$, where $\theta_{2,d^{}_{2}, d^{\prime}_{2}}$'s are unknown basis coefficients. In practice, the integration term $\int X_{i3}(t)\beta_3(t,y)dt$ is approximated by Riemann integration $\int X_{i3}(t)\beta_3(t,y) dt  \approx \sum_{j=1}^{m} X_{i3}(t_j)\beta_3(t_j,y) (t_{j+1}-t_{j})$ but other numerical approximation scheme can be also used. 

Define $Z_i(y)=\mathbbm{1}(Y_i \leq y)$ for $y \in \mathbbm{R}$. 
In practice for each $y$ in a fine grid, we view $Z_i(y)$ as a binary-valued random functional variable. It follows that the model \eqref{eq: complex model} can be written equivalently as a generalized function-on-function regression model through relating the `artificial' binary functional response $Z_i(y)$ and the mixed covariates $\mathbf{X}_{i1}, X_{i2}, X_{i3}(\cdot)$.
This model can be fitted by using for example the ideas of \cite{scheipl2015generalized} which we briefly summarize next. 

Model \eqref{eq: complex model} can be represented as the following generalized additive model,
\begin{align}
	&E[Z_{i}(y)|\mathbf{X}_{i1}, X_{i2}, X_{i3}(\cdot)] = g^{-1}\left\{ \eta_i(y) \right\}; \label{eq: additive model}\\
	&\eta_i(y) =
	\sum_{d_0=1}^{\kappa_0} B_{0,d_0}(y) \theta_{0,d_0}  + \mathbf{X}_{i1}^T\boldsymbol{\beta_1} + X_{i2} \sum_{d_1=1}^{\kappa_1} B_{1,d_1}(y) \theta_{1,d_1} \nonumber \\
	& \qquad\qquad\qquad +  \sum_{j=1}^{m} (t_{j+1}-t_{j}) X_{i3}(t_j) \sum_{d^{}_{2}=1}^{\kappa_{2,t}} \sum_{d^{\prime}_{2}=1}^{\kappa_{2,y}} B^{t}_{2,d^{}_{2}}(t_j)B^{y}_{2,d^{\prime}_{2}}(y) \theta_{2,d^{}_{2}, d^{\prime}_{2}} \nonumber
\end{align}
For convenience, we use the notation $B_{X_{i2},d_1}(y) = X_{i2} B_{1,d_1}(y)$, $\widetilde{X}_{i3}(t_j) = (t_{j+1}-t_{j}) X_{i3}(t_j)$. We let $\mathbf{B}_0(y) = \left\{B_{0,1}(y),\cdots, B_{0,\kappa_0}(y) \right\}^T$,\\ $\mathbf{B}_{i1}(y) = \left\{B_{X_{i2},1}(y),\cdots, B_{X_{i2},\kappa_1}(y) \right\}^T$, $\mathbf{B}_{2,t}(t) = \left\{B_{2,1}^t(t),\cdots, B_{2,\kappa_{2,t}}^t (t) \right\}^T$, $\mathbf{B}_{2,y}(y) = \left\{B_{2,1}^y(y),\cdots, B_{2,\kappa_{2,y}}^y (y) \right\}^T$, $\boldsymbol{\theta}_0 = \left\{\theta_{0,1},\cdots, \theta_{0,\kappa_0} \right\}^T$, $\boldsymbol{\theta}_1 = \left\{\theta_{1,1},\cdots, \theta_{1,\kappa_1} \right\}^T$,
%$\boldsymbol{\theta}_0 = \left\{\theta_{0,1},\cdots, \theta_{0,\kappa_0}(y) \right\}^T$, $\boldsymbol{\theta}_1 = \left\{\theta_{1,1},\cdots, \theta_{1,\kappa_1}(y) \right\}^T$
and $\boldsymbol{\Theta}_2 =[\theta_{2,d^{}_{2},d^{\prime}_{2}}]_{1 \leq d^{}_{2} \leq \kappa_{2,t}, 1 \leq d^{\prime}_{2} \leq \kappa_{2,y}}$ is a coefficient matrix. Then $\beta_3(t,y) = \left\{\mathbf{B}_{2,t}(t) \otimes \mathbf{B}_{2,y}(y)\right\}^T\boldsymbol{\theta}_2$, where $\boldsymbol{\theta}_2$ is the vectorization of $\boldsymbol{\Theta}_2$. We let $\widetilde{\mathbf{X}}_{i3}(\mathbf{t}) = \left\{\widetilde{X}_{i3}(t_1), \cdots, \widetilde{X}_{i3}(t_m)\right\}^T \in \mathbbm{R}^m$, $\mathbf{B}_2(\mathbf{t},y) = \left\{ \mathbf{B}_{2,t}(t_1) \otimes \mathbf{B}_{2,y}(y), \cdots, \mathbf{B}_{2,t}(t_m) \otimes \mathbf{B}_{2,y}(y)\right\}^T \in \mathbbm{R}^{m \times \kappa_{2,t}\kappa_{2,y}}$, and $\mathbf{B}_{i2}(y) = \mathbf{B}_2(\mathbf{t},y)^T \widetilde{\mathbf{X}}_{i3}(\mathbf{t})$.
Now model (\ref{eq: additive model}) can be written as 
\begin{eqnarray}
	E[Z_{i}(y)|\mathbf{X}_{i1}, X_{i2}, X_{i3}(\cdot)] = g^{-1}\left\{\mathbf{X}_{i1}^T\boldsymbol{\beta_1} +  \mathbf{B}_0(y)^T\boldsymbol{\theta}_0 + \mathbf{B}_{i1}(y)^T\boldsymbol{\theta}_1 +  \mathbf{B}_{i2}(y)^T\boldsymbol{\theta}_2\right\}.
\end{eqnarray}
The general idea is to set the bases dimensions $\kappa_0$, $\kappa_1$, $\kappa_{2,t}$ and $\kappa_{2,y}$ to be sufficiently large to capture the complexity of the coefficient functions and control the smoothness of the estimator through some roughness penalties. %$P_0(\boldsymbol{\theta}_0)$, $P_1(\boldsymbol{\theta}_1)$ and $P_2(\boldsymbol{\theta}_2)$.
This approach of using roughness penalties has been widely used; see, for example, \citet{eilers1996flexible, ruppert2002selecting, wood2003thin, wood2006} among many others.

It is important to emphasize that even in the case of a single functional covariate, our methodology differs from \citep{chen2012conditional} in two directions:
1) our proposed method is based on modeling the unknown smooth coefficient functions using pre-specified basis function expansion and using penalties to control their roughness. In contrast, CM uses data-driven basis, chooses the number of basis functions through the percentage of explained variance (PVE) of the functional predictors. This key difference allows our method to accommodate covariates of different types as well as non-linear effects.
2) Our estimation approach is based on a single step penalized function-on-function regression while CM uses pointwise estimation based on functional principal component bases and thus requires fitting multiple generalized regressions. This nice feature leads to an computational advantage.

\section{Estimation}\label{sec: estimation}
\subsection{Estimation via penalized log-likelihood} \label{subsec: pen-loglike}
Let $\{y_{\ell}: \ell = 1,\ldots, L\}$ be a set of equally spaced points in the range of the response variable, $Y_i$'s.
Conditioning on $\left\{\mathbf{X}_{i1}, X_{i2}, X_{i3}(\cdot)\right\}$, we model $Z_{i}(y_{\ell})$ as independently distributed Bernoulli variables with mean $\mu_{i}(y_{\ell})$,
where $g\{\mu_{i}(y_{\ell})\} = \eta_{i}(y_{\ell})$. The coefficients $\boldsymbol{\beta}_1$, $\boldsymbol{\theta}_0$, $\boldsymbol{\theta}_1$ and $\boldsymbol{\theta}_2$, are estimated by minimizing the penalized log-likelihood criterion,
\begin{equation}\label{eq: objective function}
	-2\mathcal{L}\left(\boldsymbol{\beta}_1, \boldsymbol{\theta}_0,\boldsymbol{\theta}_1, \boldsymbol{\theta}_2|\{Z_i(y_{\ell}\right): \forall i, \ell \} ) + \lambda_0P_0(\boldsymbol{\theta}_0) + \lambda_1P_1(\boldsymbol{\theta}_1) + \lambda_{2,t}P_{2,t}(\boldsymbol{\theta}_2) + \lambda_{2,y}P_{2,y}(\boldsymbol{\theta}_2), 
\end{equation}
where $\mathcal{L}$ is the log-likelihood function of
data $\{Z_{i}(y_{\ell}): \ell = 1, \cdots, L\}_{1 \leq i \leq n}$, $\lambda_0$, $\lambda_1$, $\lambda_{2,t}$and $\lambda_{2,y}$ are smoothing parameters, which control the balance between the model fit and its complexity, $P_0(\cdot)$, $P_1(\cdot)$, $P_{2,t}(\cdot)$ and $P_{2,y}(\cdot)$ are all penalties. 

There are several choices to define the penalty matrix in nonparametric regression, see \citet{eilers1996flexible, wood2006}.
We use quadratic penalties which penalize the size of the curvature of the estimated coefficient functions. 
Let $P_0(\boldsymbol{\theta_0}) =\int \{\partial^2 \beta_0(y) / \partial y^2 \}^2 dy = \boldsymbol{\theta}^T_0 \mathbf{D}_0 \boldsymbol{\theta}_0$, where $\mathbf{D}_0$ is of dimension $\kappa_0 \times \kappa_0$ with its $(s,s')$ element equal to $\int \{\partial^2 B_{0,s}(y) / \partial y^2\}\{\partial^2 B_{0,s'}(y) / \partial y^2 \}dy$. 
Similarly, $P_1(\boldsymbol{\theta}_1) = \int \{\partial^2 \beta_2(y) / \partial y^2 \}^2 dy = \boldsymbol{\theta}^T_1 \mathbf{D}_1 \boldsymbol{\theta}_1$, where $\mathbf{D}_1$ is of dimension $\kappa_1 \times \kappa_1$ with its $(s,s')$ element equal to $\int \{\partial^2 B_{1,s}(y) / \partial y^2\}\{\partial^2 B_{1,s'}(y) / \partial y^2 \}dy$.
As $\beta_3(\cdot,\cdot)$ is a bivariate function, the choice of penalty implies penalizing the size of curvature in each direction respectively: 
$P_{2,t}(\boldsymbol{\theta_2}) = \int \int \{\partial^2 \beta_3(y,t) / \partial t^2 \}^2 dy dt = \boldsymbol{\theta}^T_2 \mathbf{D}_{2,t} \boldsymbol{\theta}_2$, where $\mathbf{D}_{2,t} = \mathbf{P}_{2,t} \otimes \mathbf{I}_{\kappa_{2,y}}$ is of dimension $\kappa_{2,y}\kappa_{2,t} \times \kappa_{2,y}\kappa_{2,t}$ with the $(s,s')$ element of $\mathbf{P}_{2,t}$ equal to $\int \{\partial^2 B^{t}_{2,s}(t) / \partial t^2\} \allowbreak \{\partial^2 B^{t}_{2,s'}(t) / \partial t^2\} dt$ for some orthonormal spline bases.
Similarly, $P_{2,y}(\boldsymbol{\theta_2}) = \int \int \{\partial^2 \beta_3(y,t) / \partial y^2 \}^2 dy dt =\boldsymbol{\theta}^T_2 \mathbf{D}_{2,y} \boldsymbol{\theta}_2$, where $\mathbf{D}_{2,y} = \mathbf{I}_{\kappa_{2,t}} \otimes \mathbf{P}_{2,y} $ is of dimension $\kappa_{2,y}\kappa_{2,t} \times \kappa_{2,y}\kappa_{2,t}$ with the $(s,s')$ element of $\mathbf{P}_{2,y}$ equal to $\int \{\partial^2 B^{y}_{2,s}(y) / \partial y^2 \} \{  \partial^2 \allowbreak B^{y}_{2,s'}(y) / \partial y^2\} dy $.

%In the following, we detail the estimation algorithm.  
The criterion (\ref{eq: objective function}) can be viewed as a penalized quasi-likelihood (PQL) of the corresponding generalized linear mixed model
\begin{eqnarray} \label{eq: mixed model scalar covariate}
	Z_{i}(y_{\ell})|\boldsymbol{\beta}_1,\boldsymbol{\theta}_0,\boldsymbol{\theta}_1, \boldsymbol{\theta}_2 &\sim& \text{Bernoulli}(\mu_{i}(y_{\ell})), \quad \ell = 1, \ldots, L;\\ \quad \boldsymbol{\theta}_0 &\sim& N\Big( \textbf{0}, \lambda_0^{-1} \mathbf{D}^{-}_0 \Big);
	\quad \boldsymbol{\theta}_1 \sim N\Big( \textbf{0}, \lambda_1^{-1} \mathbf{D}^{-}_1 \Big);\nonumber\\
	\quad \boldsymbol{\theta}_{2} &\sim& N\Big( \textbf{0}, \mathbf{D}^{-}_{2} \Big), \nonumber
	%\quad \boldsymbol{\theta}_{2,y} \sim N\Big( \textbf{0}, \lambda_{2,y}^{-1} \mathbf{D}^{-}_{2,y} \Big),\nonumber
\end{eqnarray}
where $\mathbf{D}^{-}_0$ is the generalized inverse matrix of $\mathbf{D}_0$; $\mathbf{D}^{-}_1$ and $\mathbf{D}^{-}_{2} = (\lambda_{2, t} \mathbf{D}_{2, t} + \lambda_{2, y} \mathbf{D}_{2, y})^-$ are defined similarly. \citet{wood2006} discusses an alternative way to deal with the rank-deficient matrices in the context of restricted maximum likelihood (REML) estimation. 
Here we do not account for the dependence over $y$, see \cite{scheipl2015generalized} for a general formulation.
See also \citet{ivanescu2014penalized} who uses the mixed model representation of a similar regression model to (\ref{eq: mixed model scalar covariate}), but with a Gaussian functional response. The smoothing parameters are estimated using REML.

\subsection{Extension to nonlinear model}\label{subsec: nonlinear}

One advantage of the proposed framework is that it can be easily extended to allow for more flexible effects, i.e., extending the ideas to accommodate multiple covariates, scalar or functional, and varied types of effects.
In particular, the smooth effect $X_{i2}\beta_2(y)$ can be replaced by $h_{1}(X_{i2},y)$, and $\int X_{i3}(t) \beta_{3}(t,y)dt$ by $\int h_{2}\{X_{i3}(t),t,y\}dt$, where 
$h_1(\cdot, \cdot)$ and $h_2(\cdot, \cdot, \cdot)$ are unknown bivariate and trivariate smooth functions, respectively; see \cite{scheipl2015generalized} and \cite{kim2018additive}. These changes require little additional computational effort. The modeling and estimation follow roughly similar ideas as \cite{scheipl2015generalized}.
%In particular, the smooth effect $X_{i2}\beta_2(y)$ can be replaced by $h_{1}(X_{i2},y)$, and the linear effect $\int X_{i3}(t) \beta_{3}(t,y)dt$ by $\int h_{2}\{X_{i3}(t),t,y\}dt$, where $h_1(\cdot)$ and $h_2(\cdot, \cdot, \cdot)$ are unknown univariate and trivariate smooth functions, respectively; these changes require little additional computational burden, see \cite{scheipl2015generalized}. 
%Furthermore the proposed method can be easily extended to relax the linearity assumption and allow more flexible model structures. Instead of a functional linear model such as $F_{Y|X}(y) = g^{-1}\big\{ \beta_0(y) + X_1\beta_1(y) + \int X_2(t)\beta_2(t,y)dt \big\}$, we can model the conditional distribution as $F_{Y|X}(y) = g^{-1}\big\{ \beta_0(y) + h_1(X_1) + \int h_2(X_2(t), t, y)dt \big\}$, where $h_1(\cdot)$ and $h_2(\cdot, \cdot, \cdot)$ are unknown univariate and trivariate smooth functions respectively.
%This nonlinear model can still be implemented using the \verb|pffr| function with \verb|sff| in place of \verb|ff|.
We consider the nonlinear model in the simulation study for the case of having a scalar covariate only, i.e. $h_1(X_i, y)$,
%for the case of having a scalar covariate only 
and the corresponding results are presented in Section \ref{sec: sim-sca} of the Supplementary materials. The results show excellent prediction performance compared to the competitive nonlinear quantile regression method, namely Constrained B-Spline Smoothing (COBS) \citep{Ng2007cobs}.

\subsection{Extension to sparse and noisy functional covariates}\label{subsec: Extension to sparse and noisy functional covariates}

In practice the functional covariates are often observed at irregular times across the units and the measurements are possibly corrupted by noises. In such case, one needs to first smooth and de-noise the trajectories before fitting. When the sampling design of the functional covariate is dense, the common approach is to smooth each trajectory using splines or local polynomial smoothing, as proposed in \citet{ramsay2006functional} and \citet{zhang2007statistical}. When the design is sparse, the smoothing can be done by pooling all the subjects and following the PACE method proposed in \citet{yao2005functional}. As recovering the trajectories has been extensively discussed in the literatures, we do not review the procedures here. Instead, we discuss some available computing resources that can be used to fit these methods. In our numerical study, we used \texttt{fpca.sc} function in the \texttt{refund} \texttt{R} package \citep{Rrefund} for recovering the latent trajectories, irrespective of a sampling design (dense or sparse).  
Alternatively, one can use \texttt{fpca.face} \citep{xiao2016fast} in \texttt{refund} for regular dense design and \texttt{face.sparse} \citep{xiao2016face} in the \texttt{R} package \texttt{face} \citep{Xiao:16c} for irregular sparse design. Once the latent trajectories are estimated, they can be used in the fitting criterion (\ref{eq: objective function}).

\subsection{Estimation of conditional quantile}
Let $\widehat{\boldsymbol{\beta}}_1$, $\widehat{\boldsymbol{\theta}}_0$, $\widehat{\boldsymbol{\theta}}_1$, and $\widehat{\boldsymbol{\theta}}_2$ be parameter estimates in (\ref{eq: mixed model scalar covariate}). It follows that the estimated distribution function  $\widehat{F}_{Y_i|\mathbf{X}_{i1}, X_{i2}, X_{i3}(\cdot)}(y)$ can be obtained by plugging in the estimated coefficients. The $\tau$th conditional quantile is estimated  by inverting the estimated distribution, i.e., $\widehat{Q}_{Y_i|\mathbf{X}_{i1}, X_{i2}, X_{i3}(\cdot)}(\tau) = \text{inf}\{y: \widehat{F}_{Y_i|\mathbf{X}_{i1}, X_{i2}, X_{i3}(\cdot)}(y) \geq \tau \}$. %This approach relates the $\tau$th level quantile of the response in a nonlinear manner to the covariate. 
The estimated distribution function is not a monotonic function yet. In practice we suggest to first apply a monotonization method as described in the Section \ref{subsec: monotonization}, and then estimate the conditional quantiles by inverting the resulting estimated distribution.

%\begin{comment}	
\subsection{Monotonization and implementation}\label{subsec: monotonization}

While a conditional quantile function is nondecreasing, the resulting estimated quantiles may not be.  %\citet{chernozhukov2009improving} showed that in this way the monotonized estimator gives the same or better fit than the original estimator. 
Two approaches are widely used: one is to monotonize the estimated conditional distribution function, and the other is to monotonize the estimated conditional quantile function. We choose the former as $\widehat{F}_{Y_i|\mathbf{X}_{i1}, X_{i2}, X_{i3}(\cdot)}(y)$ is readily available at dense grid points $y_{\ell}$'s. We use an isotonic regression model  \citep{barlow1972statistical} for monotonization, which imposes an order restriction; this is done by using the \texttt{R} function \texttt{isoreg}. 
Other monotonization approaches include  \citet{chernozhukov2009improving}, which was employed in \citet{kato2012estimation}.

%\end{comment}	

%\subsection{Implementation}\label{subsec:implementation}
Our approach is implemented by first creating an artificial binary response and then fitting a penalized function-on-function regression model and using the logit link function.
Fitting models in (\ref{eq: objective function}) can be done by extending the ideas of \citet{ivanescu2014penalized} for Gaussian functional response; the extension of the model to the non-Gaussian functional response has recently been studied and implemented by \citet{scheipl2015generalized} as the \texttt{pffr} function in \texttt{refund} package \citep{Rrefund}.

\section{Simulation study} \label{sec: simulation}
\subsection{Simulation setting}
In this section we evaluate the empirical performance of the proposed method. We present results for the case when we have both functional and scalar covariates; additional results when there is only a single scalar or a single functional covariate are discussed in the Supplementary materials, Section \ref{sec: sim-add}. %We adapt the simulation settings of \citet{chen2012conditional} for the cases that involve a functional covariate.

Suppose the observed data for the $i$th subject are $[Y_i, X_{1i}, \{(W_{i1},t_{i1}), \cdots, \allowbreak (W_{i m_i},t_{i m_i})\}]$, $t_{ij} \in [0,10]$, where $X_{1i} \overset{i.i.d} \sim Unif(-16,16)$, $W_{ij} = X_{2i}(t_{ij}) + \epsilon_{ij}$ for $1 \leq i \leq n, 1 \leq j \leq m_i$. Let $X_{2i}(t_{ij})= \mu(t_{ij}) + \sum_{k=1}^4 \xi_{ik} \phi_k(t_{ij}) + \epsilon_{ij}$, where $\mu(t) = t + \sin(t)$, $\phi_k (t) = \cos\{ (k + 1) \pi t / 10 \} / \sqrt{5}$ for odd values of $k$, $\phi_k (t) = \sin\{ k \pi t / 10 \} / \sqrt{5}$ for even values of $k$, $\xi_{ik} \stackrel{iid}{\sim} N(0, \lambda_k)$, $(\lambda_1,\lambda_2,\lambda_3,\lambda_4) = \{ 16, 9, 7.56, 5.06 \}$, and $\epsilon_{ij} \stackrel{iid}{\sim} N(0, \sigma^2_{\epsilon})$.
%Three noise levels are considered: %$\sigma^2=0$ (no noise);
%low ($\sigma_{\epsilon}=0.50$), moderate ($\sigma_{\epsilon}=4.33$), and high ($\sigma_{\epsilon}=6.13$). The three levels are such that the signal to noise ratio (SNR), which are calculated as $SNR = \sqrt{\sum_{k=1}^4\lambda_k} / \sigma_{\epsilon}$, are equal to $SNR = 150$, $2$, and $1$, respectively.
We assume three cases for generating response $Y_i$:
\begin{enumerate}
	\item[(i)] Gaussian: $Y_i|X_{1i},X_{2i}(\cdot) \sim N(2\int X_{2i}(t) \beta(t) dt + 2 X_{1i}, 5^2)$; this corresponds to the quantile regression model $Q_{Y|X_{1},X_{2}(\cdot)}(\tau) = 2 \int X_{2i}(t) \beta(t) dt + 2 X_{1i} + 5 \Phi^{-1}(\tau)$, where $\Phi(\cdot)$ is the distribution function of the standard normal;
	\item[(ii)] Mixture of Gaussians: \\ $Y_i|X_{1i},X_{2i}(\cdot) \sim 0.5N(\int X_{2i}(t) \beta(t) dt+ X_{1i}, 1^2) +$ $0.5 $ $N(3\int X_{2i}(t) \beta(t) dt$ $+ 3 X_{1i}, 4^2)$, where the true quantiles can be approximated numerically by using \texttt{qnorMix} function in the \texttt{R} package \texttt{norMix};
	%And $\beta(t) = \sum_{k=1}^4 \beta_{k} \phi_k(t)$. 
	\item[(iii)]  Gaussian with heterogeneous error: $Y_i|X_{1i},X_{2i}(\cdot) \sim N(2\int X_{2i}(t) \beta(t) dt + 2 X_{1i}, 5^2 \int X_{2i}(t)^2 dt / \sum_{k=1}^4 \lambda_k)$; the true quantiles are given by $Q_{Y|X_{1},X_{2}(\cdot)}(\tau) = 2 \int X_{2i}(t) \beta(t) dt + 2 X_{1i} + 5 \sqrt{\int X_{2i}(t)^2 dt / \sum_{k=1}^4 \lambda_k} \Phi^{-1}(\tau)$.
\end{enumerate}
Let $\beta(t) = \sum_{k=1}^4 \beta_{k} \phi_k(t)$, where $\beta_k =1$ for $k = 1, \cdots, 4$.

For each case, we use different combinations of signal to noise ratio (SNR), sample size and sampling designs to generate $500$ simulated datasets.
We define SNR as $\sqrt{\sum_{k=1}^4\lambda_k} / \sigma_{\epsilon}$, and consider five levels of noise: %$\sigma^2=0$ (no noise);
$\textnormal{SNR} = \{150, 10, 5, 2, 1\}$.
Two levels of sample size are $n=100$ and $n=1000$. 
Two sampling designs are considered: (i) \textit{sparse design}, where $\{t_{ij}: j=1, \ldots, m\} $ are $m=15$ randomly selected points from a set of $30$ equi-spaced grids in $[0,10]$; and (ii) \textit{dense design}, where the sampling points $\{ t_{ij} = t_j: j=1, \ldots, m\} $ are $m=30$ equi-spaced time points in $[0,10]$.

The performance is evaluated on a test set of $100$ subjects, for which we have $\{X_{1i^{*}}, (W_{i^{*}j}, t_{i^{*}j}), j = 1, \ldots, m\}$ available, in terms of mean absolute error (MAE) for quantile levels $\tau = 0.05, 0.1, 0.25,$ and $0.5$,
\[
\textnormal{MAE}(\tau) = \frac{1}{100}\sum_{i^{*}=1}^{100}|\widehat{Q}_{i^{*}}(\tau) - Q_{i^{*}}(\tau)|.
\]

\subsection{Competing methods}
%we use the \texttt{pffr} function \citep{ivanescu2014penalized,scheipl2015generalized} in the \texttt{refund} package \citep{Rrefund} in \texttt{R} \citep{R} for binomial responses, denoted by Joint QR.
We denote the proposed method by Joint QR to emphasize the single step estimation approach.
We compare our method with two alternative approaches: (1) a variant of our proposed approach using pointwise estimation, denoted by Pointwise QR. 
This approach consists of fitting multiple regression models with binomial link function as implemented by the penalized functional regression \texttt{pfr}, developed by \citet{goldsmith2012penalized}, of the \texttt{refund} package for generalized scalar responses. 
(2) A modified version of the CM method, denoted by Mod CM, that we developed to account for additional scalar covariates, and which fits multiple generalized linear models with scalar covariates and fPC scores as predictors.
% \rev{the number of principal components are selected according to $PVE = 0.99$};
(3) A linear quantile regression approach using the quantile loss function and the partial quantile regression bases for functional covariates, proposed by \citet{yu2015partial} and denoted by PQR. 
Notice that although the formulation of the first two methods implicitly account for a varying effect of the covariates on the response distribution, they do not ensure that this effect is smooth.
The third approach can only estimate a specific quantile rather than the entire conditional distribution.
Note that all the competing methods are monotonized for a fair comparison.

The {\tt R} function \texttt{pfr} can incorporate both scalar/vector and functional predictors by adopting a mixed effects model framework. The functional covariates are pre-smoothed by fPC analysis \citep{yao2005functional}; %pre-smoothing the functional covariates before fitting the regression model has been also considered by \citet{goldsmith2012penalized} and \citet{ivanescu2014penalized}.
Throughout the simulation study we fix PVE as $0.95$ for fPC analysis to determine the mumber of principal components and use REML to select the smoothing parameters for our proposed methods. Other basis settings are set to their default values.
We use 100 equally-distanced points between the minimum and maximum of the observed $Y_i$'s to set the grid $\left\{y_{\ell}: \ell = 1, \cdots, L \right\}$ for the conditional distribution function.

%For the PQR method we  set the number of basis functions $K = 5$.

\subsection{Simulation results}

Tables \ref{tab:sim1MAE1} and \ref{tab:sim1MAE2} show the accuracy of the quantile estimation for the two cases (normal and mixture) when the functional covariate is observed sparsely and the sample size is $n=100$ (Table \ref{tab:sim1MAE1}) and  $n=1000$ (Table \ref{tab:sim1MAE2}). 
Table \ref{tab:sim1MAE3} presents the estimation accuracy for the case of heteroskedasticity with sparsely observed functional covariates.
The results based on dense sampling design show similar patterns and thus are relegated to the supplement, see Section \ref{sec: sim-den}.
The comparison of running times is presented in Table \ref{tab:sim1time}.

For the case when the response is Gaussian, Tables \ref{tab:sim1MAE1} and \ref{tab:sim1MAE2} suggest that the Joint QR typically outperforms its competitors especially for lower quantile levels ($\tau = 0.05$ and $\tau =0.1$). For very small noise level (SNR$=150$), PQR performs the best, followed closely by the proposed Joint QR. The variant Pointwise QR, which has a poorer performance, is generally better than the modified CM approach. As expected, as the sample size increases ($n = 1000$), all the accuracy results improve; the proposed Joint QR continues to yield most accurate quantiles for the low quantile levels. For mixture of Gaussians, the results are somewhat similar. The accuracy of the quantile estimators with the Pointwise QR improves greatly; in fact the Joint QR and Pointwise QR outperform the other approaches for quantile levels $\tau = 0.05, 0.1, 0.25$ irrespective of the SNR. Finally, Table \ref{tab:sim1MAE3} shows that the results for Gaussian with heterogeneous error are close to those for the case of Gaussian. Again, the proposed method has competitive performance in terms of estimation accuracy.

Table \ref{tab:sim1time} compares the three methods that involve estimating the conditional distribution in terms of the running time required for fitting. 
The times are reported based on a computer with a $2.3$ GHz CPU and $8$ GB of RAM. 
Not surprisingly by fitting the model a single time, Joint QR is the fastest, in some cases being order of magnitude faster than the rest. Pointwise QR can be up to twice as fast as Mod CM.

%In addition to the three methods discussed above, we compare the proposed method with the following approaches by either ignoring or modifying covariates of one type: $(1)$ linear quantile regression (LQR), as implemented by the \texttt{rq} function of the \texttt{quantreg} \citep{koenker2005quantile} package in \texttt{R} by ignoring the functional covariate; $(2)$ 	LQR using in addition a summary of functional covariate, mean of the functional covariate $\bar{X}_{2i} = \int X_{2i}(t)dt$ denoted by LQR(M), note that this competitor makes sense only for dense design;		$(3)$ the functional quantile regression proposed in \cite{chen2012conditional} (CM) by ignoring the scalar covariate, and as implemented in the \texttt{MATLAB} toolbox \texttt{PACE}. The results for these methods are presented in Tables \ref{suptab:sim1MAE1} and \ref{suptab:sim1MAE2} of the Supplementary Materials, Section \ref{sec: sim-add-mixed}. As expected, these approaches have a poorer MAE performance as they either ignore one of the covariates or do not account properly for the functional covariate.	

For completeness, we also compare our proposed method to the appropriate competitive methods for the cases $(1)$ when there is a single scalar covariate and $(2)$ when there is a single functional covariate. The Supplementary materials, Section \ref{sec: sim-sca} discusses the former case and compares Joint QR and Pointwise QR with the linear quantile regression (LQR) \citep{koenker1978regression}, implemented by \texttt{rq} function in the \texttt{R} package \texttt{quantreg}, and the constrained B-splines nonparametric regression quantiles (COBS), implemented by the \texttt{cobs} function in the \texttt{R} package \texttt{COBS} \citep{Ng2007cobs}, in an extensive simulation experiment that involves both linear quantile settings and nonlinear quantile settings. Overall the results show that the proposed methods have similar behavior as LQR, see Table \ref{tab:sim2MAE}. Furthermore we consider the proposed method and its variant with nonlinear modeling of the conditional distribution as discussed in Section \ref{subsec: nonlinear}, which we denote with Joint QR (NL) for joint fitting and Pointwise QR (NL) for pointwise fitting. Nonlinear versions of the proposed methods have an excellent MAE performance, which is comparable to or better than that of the COBS method.

Finally, Section \ref{sec: sim-fun} in the Supplementary materials discusses the simulation study for the case of having a single functional covariate and compares the proposed methods with CM in terms of MAE as well as computation time; see results displayed in Tables \ref{tab: sim3MAE} and \ref{tab:sim3time}. The results show that the proposed Joint QR is comparable to CM in terms of the prediction accuracy and has less computation time. In our simulation study we also consider the joint fitting of the model by treating the binary response as normal and use \texttt{pffr} \citep{ivanescu2014penalized} with Gaussian link, denoted by Joint QR (G).

\begin{table}[htbp]
	\caption{Average MAE (standard error in parentheses) of the predicted $\tau$-level quantile for the case of having a scalar covariate and a sparsely observed functional covariate. Sample size $n=100$.}
	\centering
	\scalebox{0.75}
	{
		\begin{tabular}{ccccccc}
			\hline\hline
			Distribution &  SNR & Method & $\tau=0.05$ & $\tau=0.1$ & $\tau=0.25$ & $\tau=0.5$ \\
			\hline\hline

			\multirow{4}{*}{Normal}&\multirow{4}{*}{150}& Joint QR & 3.67  (0.03) & 3.53  (0.03) & 3.30  (0.02) & 3.17  (0.02) \\ 
			\multirow{4}{*}{}&\multirow{4}{*}{}& Pointwise QR & 4.96  (0.03) & 4.61  (0.03) & 4.22  (0.02) & 4.18  (0.02) \\ 
			\multirow{4}{*}{}&\multirow{4}{*}{}& Mod CM & 6.04 (0.03) & 5.81 (0.03) & 5.55 (0.03) & 5.14 (0.03) \\
			\multirow{4}{*}{}&\multirow{4}{*}{}& PQR & \textbf{3.17}  (0.04) & \textbf{2.71} (0.03) & \textbf{2.31}  (0.02) & \textbf{2.16 } (0.02) \\   [0.3cm]
			
			\multirow{4}{*}{Normal}&\multirow{4}{*}{10}& Joint QR & \textbf{6.32}  (0.03) & \textbf{6.00} (0.03) & 5.76  (0.02) & 5.73  (0.02) \\
			\multirow{4}{*}{}&\multirow{4}{*}{}& Pointwise QR & 7.44  (0.04) & 6.85  (0.03) & 6.39  (0.03) & 6.28  (0.03) \\ 
			\multirow{4}{*}{}&\multirow{4}{*}{}& Mod CM & 8.20 (0.04) & 8.10 (0.04) & 8.04 (0.04) & 8.01  (0.04) \\
			\multirow{4}{*}{}&\multirow{4}{*}{}& PQR  & 6.82  (0.05) & 6.11  (0.04) & \textbf{5.34}  (0.03) & \textbf{5.09}  (0.02) \\    [0.3cm]
			
			\multirow{4}{*}{Normal}&\multirow{4}{*}{5}& Joint QR & \textbf{7.84}  (0.04) & \textbf{7.34}  (0.03) & 6.93  (0.03) & 6.84  (0.03) \\ 
			\multirow{4}{*}{}&\multirow{4}{*}{}& Pointwise QR & 8.91  (0.04) & 8.12  (0.04) & 7.45  (0.03) & 7.26  (0.03) \\ 
			\multirow{4}{*}{}&\multirow{4}{*}{}& Mod CM & 9.34 (0.04) & 9.23 (0.04) & 9.14 (0.05) & 9.06  (0.05) \\
			\multirow{4}{*}{}&\multirow{4}{*}{}& PQR  & 8.68  (0.06) & 7.81  (0.05) & \textbf{6.74}  (0.03) & \textbf{6.34}  (0.02) \\    [0.3cm]

			\multirow{4}{*}{Normal}&\multirow{4}{*}{2}& Joint QR & \textbf{10.05} (0.05) & \textbf{9.22}  (0.04) & \textbf{8.47}  (0.03) & 8.28  (0.03) \\ 
			\multirow{4}{*}{}&\multirow{4}{*}{}& Pointwise QR &  10.91  (0.06) & 9.86  (0.04) & 8.87  (0.04) & 8.54  (0.03) \\ 
			\multirow{4}{*}{}&\multirow{4}{*}{}& Mod CM & 10.85 (0.05) & 10.55 (0.05) & 10.34 (0.06) & 10.21 (0.06) \\
			\multirow{4}{*}{}&\multirow{4}{*}{}& PQR & 11.21  (0.08) & 10.03  (0.06) & 8.56  (0.04) & \textbf{7.96 } (0.03) \\   [0.3cm]
			
			\multirow{4}{*}{Normal}&\multirow{4}{*}{1}& Joint QR & \textbf{11.50}  (0.06) & \textbf{10.41}  (0.05) & \textbf{9.40 } (0.04) & 9.11  (0.03) \\ 
			\multirow{4}{*}{}&\multirow{4}{*}{}& Pointwise QR & 12.12  (0.06) & 10.88  (0.05) & 9.70  (0.04) & 9.30  (0.03) \\ 
			\multirow{4}{*}{}&\multirow{4}{*}{}& Mod CM & 11.95 (0.06) & 11.46 (0.06) & 11.07 (0.06) & 11.05 (0.07 )\\
			\multirow{4}{*}{}&\multirow{4}{*}{}& PQR &  12.82  (0.08) & 11.38  (0.06) & 9.60  (0.04) & \textbf{8.86 } (0.03) \\  [0.1cm]
			\hline \\  [-0.3cm]

			\multirow{4}{*}{Mixture}&\multirow{4}{*}{150}& Joint QR & \textbf{6.92}  (0.06) & \textbf{6.23}  (0.06) & \textbf{6.16}  (0.06) & \textbf{4.81 } (0.06) \\ 
			\multirow{4}{*}{}&\multirow{4}{*}{}& Pointwise QR & 8.10  (0.08) & 6.80  (0.06) & 6.66  (0.06) & 5.25  (0.06) \\ 
			\multirow{4}{*}{}&\multirow{4}{*}{}& Mod CM & 9.18 (0.07) & 8.99 (0.07) & 8.93 (0.07) & 7.90 (0.07)\\ 
			\multirow{4}{*}{}&\multirow{4}{*}{}& PQR & 8.43  (0.06) & 7.18  (0.04) & 6.22  (0.04) & 5.48  (0.14) \\  [0.3cm]
			
			\multirow{4}{*}{Mixture}&\multirow{4}{*}{10}& Joint QR & \textbf{9.02} (0.06) & \textbf{7.95}  (0.05) & 7.85 (0.05) & 6.19  (0.06) \\
			\multirow{4}{*}{}&\multirow{4}{*}{}& Pointwise QR & 10.11  (0.07) & 8.52  (0.05) & 7.95  (0.05) & 6.42  (0.06) \\ 
			\multirow{4}{*}{}&\multirow{4}{*}{}& Mod CM & 11.33 (0.07) & 10.95 (0.07) & 10.79 (0.07) & 9.80 (0.08)\\ 
			\multirow{4}{*}{}&\multirow{4}{*}{}& PQR & 10.72  (0.08) & 8.99  (0.05) & \textbf{7.63}  (0.04) & \textbf{5.33}  (0.09) \\  [0.3cm]
			
			\multirow{4}{*}{Mixture}&\multirow{4}{*}{5}& Joint QR & \textbf{10.18} (0.06) & \textbf{8.91}  (0.05) & 8.53  (0.05) & 6.82  (0.05) \\
			\multirow{4}{*}{}&\multirow{4}{*}{}& Pointwise QR  & 11.18  (0.07) & 9.38  (0.05) & 8.58  (0.05) & 6.98  (0.05) \\ 
			\multirow{4}{*}{}&\multirow{4}{*}{}& Mod CM & 12.19 (0.07) & 11.75 (0.07) & 11.52 (0.07) & 10.47 (0.08) \\ 
			\multirow{4}{*}{}&\multirow{4}{*}{}& PQR & 12.12  (0.09) & 10.12 (0.06) & \textbf{8.40} (0.04) & \textbf{5.73}  (0.07) \\  [0.3cm]
			
			\multirow{4}{*}{Mixture}&\multirow{4}{*}{2}& Joint QR &  \textbf{11.93}  (0.07) & \textbf{10.26}  (0.05) & \textbf{9.46}  (0.05) & 7.61  (0.05) \\ 
			\multirow{4}{*}{}&\multirow{4}{*}{}& Pointwise QR & 12.68  (0.08) & 10.63  (0.06) & 9.51  (0.05) & 7.70  (0.05) \\ 
			\multirow{4}{*}{}&\multirow{4}{*}{}& Mod CM & 13.33 (0.08) & 12.60 (0.08) & 12.27 (0.08) & 11.08 (0.10) \\
			\multirow{4}{*}{}&\multirow{4}{*}{}& PQR & 14.16  (0.10) & 11.72  (0.06) & 9.55  (0.04) & \textbf{6.41}  (0.05) \\ [0.3cm]
			
			\multirow{4}{*}{Mixture}&\multirow{4}{*}{1}& Joint QR & \textbf{13.17}  (0.08) & \textbf{11.19}  (0.05) & \textbf{10.06}  (0.05) & 8.04 (0.05) \\ 
			\multirow{4}{*}{}&\multirow{4}{*}{}& Pointwise QR & 13.71  (0.09) & 11.44  (0.06) & 10.10  (0.05) & 8.13  (0.05) \\ 
			\multirow{4}{*}{}&\multirow{4}{*}{}& Mod CM & 14.16 (0.08) & 13.22 (0.08) & 12.71 (0.09) & 11.44 (0.11) \\
			\multirow{4}{*}{}&\multirow{4}{*}{}& PQR & 15.44  (0.11) & 12.84  (0.07) & 10.23  (0.04) &\textbf{6.89}  (0.04) \\ 
			
			\hline\hline
		\end{tabular}
	}
	\label{tab:sim1MAE1}
\end{table}

\begin{table}[htbp]
	\caption{Average MAE (standard error in parentheses) of the predicted $\tau$-level quantile for the case of having a scalar covariate and a sparsely observed functional covariate. Sample size $n=1000$.}
	\centering
	\scalebox{0.75}
	{
		\begin{tabular}{ccccccc}
			\hline\hline
			Distribution & SNR & Method & $\tau=0.05$ & $\tau=0.1$ & $\tau=0.25$ & $\tau=0.5$ \\
			\hline\hline
			
			\multirow{4}{*}{Normal}&\multirow{4}{*}{150}& Joint QR &  \textbf{1.68} (0.01) & 1.65  (0.01) & 1.62  (0.01) & 1.60  (0.01) \\
			\multirow{4}{*}{}&\multirow{4}{*}{}& Pointwise QR & 1.94  (0.01) & 1.92  (0.01) & 1.88  (0.01) & 1.81  (0.01) \\ 
			\multirow{4}{*}{}&\multirow{4}{*}{}& Mod CM & 1.93 (0.01) & 1.88 (0.01) & 1.87 (0.01) & 1.87 (0.01) \\ 
			\multirow{4}{*}{}&\multirow{4}{*}{}& PQR & 1.72 (0.01) & \textbf{1.61}  (0.01) & \textbf{1.51}  (0.01) & \textbf{1.48}  (0.01) \\ [0.3cm]
			
			\multirow{4}{*}{Normal}&\multirow{4}{*}{10}& Joint QR & \textbf{5.45}  (0.02) & \textbf{4.97}  (0.02) &\textbf{4.66}  (0.02) & 4.65  (0.02) \\
			\multirow{4}{*}{}&\multirow{4}{*}{}& Pointwise QR & 5.64  (0.02) & 5.13  (0.02) & 4.78  (0.02) & 4.75  (0.02) \\
			\multirow{4}{*}{}&\multirow{4}{*}{}& Mod CM & 5.69 (0.02) & 5.21 (0.02) & 4.87 (0.02) & 4.85 (0.02) \\
			\multirow{4}{*}{}&\multirow{4}{*}{}& PQR & 5.85  (0.02) & 5.37  (0.02) & 4.81  (0.02) & \textbf{4.60}  (0.02) \\ [0.3cm]
			
			\multirow{4}{*}{Normal}&\multirow{4}{*}{5}& Joint QR & \textbf{7.34}  (0.03) & \textbf{6.54}  (0.02) & \textbf{5.94}  (0.02) & 5.85  (0.02) \\
			\multirow{4}{*}{}&\multirow{4}{*}{}& Pointwise QR & 7.53  (0.03) & 6.69  (0.02) & 6.04  (0.02) & 5.94  (0.02) \\  
			\multirow{4}{*}{}&\multirow{4}{*}{}& Mod CM & 7.53 (0.03) & 6.77 (0.02) & 6.18 (0.02) & 6.06 (0.02) \\ 
			\multirow{4}{*}{}&\multirow{4}{*}{}& PQR & 7.84  (0.03) & 7.05  (0.02) & 6.16  (0.02) & \textbf{5.81}  (0.02) \\[0.3cm]
			
			\multirow{4}{*}{Normal}&\multirow{4}{*}{2}& Joint QR &\textbf{ 9.97}  (0.03) & \textbf{8.70}  (0.03) & \textbf{7.62 } (0.02) & 7.38  (0.02) \\ 
			\multirow{4}{*}{}&\multirow{4}{*}{}& Pointwise QR & 10.15  (0.03) & 8.85  (0.03) & 7.71  (0.02) & 7.45  (0.02) \\ 
			\multirow{4}{*}{}&\multirow{4}{*}{}& Mod CM & 10.12 (0.03) & 8.93 (0.03) & 7.87 (0.03) & 7.60  (0.03) \\ 
			\multirow{4}{*}{}&\multirow{4}{*}{}& PQR &10.55  (0.04) & 9.34  (0.03) & 7.91  (0.03) & \textbf{7.34 } (0.02) \\  [0.3cm]
			
			\multirow{4}{*}{Normal}&\multirow{4}{*}{1}& Joint QR & \textbf{11.69 } (0.04) & \textbf{10.10}  (0.03) & \textbf{8.68}  (0.03) & 8.32  (0.03) \\ 
			\multirow{4}{*}{}&\multirow{4}{*}{}& Pointwise QR & 11.88  (0.04) & 10.25  (0.04) & 8.77  (0.03) & 8.39  (0.03) \\ 
			\multirow{4}{*}{}&\multirow{4}{*}{}& Mod CM & 11.85 (0.04) & 10.37 (0.04) & 8.96 (0.03) & 8.57 (0.03) \\ 
			\multirow{4}{*}{}&\multirow{4}{*}{}& PQR & 12.29  (0.04) & 10.77  (0.04) & 9.02  (0.03) & \textbf{8.29}  (0.03) \\ [0.1cm]
			\hline \\  [-0.3cm]
			\multirow{4}{*}{Mixture}&\multirow{4}{*}{150}& Joint QR & 4.56 (0.02) & 4.44  (0.02) & 4.33  (0.03) & 3.68  (0.03) \\ 
			\multirow{4}{*}{}&\multirow{4}{*}{}& Pointwise QR & \textbf{4.34}  (0.03) & \textbf{4.24 } (0.02) & \textbf{4.20}  (0.03) & 3.66  (0.04) \\ 
			\multirow{4}{*}{}&\multirow{4}{*}{}& Mod CM &  4.68 (0.03) & 4.59 (0.02) & 4.38 (0.03) & 4.01 (0.03)\\
			\multirow{4}{*}{}&\multirow{4}{*}{}& PQR & 7.84  (0.03) & 6.45  (0.02) & 5.29  (0.02) & \textbf{3.19}  (0.01) \\  [0.3cm]
			
			\multirow{4}{*}{Mixture}&\multirow{4}{*}{10}& Joint QR & 7.56  (0.03) & 6.79  (0.02) & 6.14  (0.03) & 5.02  (0.03) \\  
			\multirow{4}{*}{}&\multirow{4}{*}{}& Pointwise QR & \textbf{7.49}  (0.03) & \textbf{6.61}  (0.02) & \textbf{5.88}  (0.03) & 5.02  (0.03) \\
			\multirow{4}{*}{}&\multirow{4}{*}{}& Mod CM & 7.67 (0.04) & 6.97 (0.03) & 6.17 (0.03) & 5.52 (0.04)\\
			\multirow{4}{*}{}&\multirow{4}{*}{}& PQR  & 10.11  (0.04) & 8.29  (0.03) & 6.71  (0.02) & \textbf{3.62}  (0.02) \\  [0.3cm]
			
			\multirow{4}{*}{Mixture}&\multirow{4}{*}{5}& Joint QR  & 9.20  (0.04) & 7.13  (0.03) & 6.94  (0.03) & 5.6  (0.03) \\ 
			\multirow{4}{*}{}&\multirow{4}{*}{}& Pointwise QR & \textbf{9.17}  (0.04) & \textbf{7.01}  (0.03) & \textbf{6.71}  (0.03) & 5.58 (0.03) \\
			\multirow{4}{*}{}&\multirow{4}{*}{}& Mod CM & 9.32 (0.04) & 7.37 (0.03) & 7.08 (0.03) & 6.18 (0.04) \\
			\multirow{4}{*}{}&\multirow{4}{*}{}& PQR & 11.46  (0.04) & 9.36  (0.03) & 7.49  (0.03) & \textbf{4.32}  (0.02) \\ [0.3cm]
			
			\multirow{4}{*}{Mixture}&\multirow{4}{*}{2}& Joint QR & \textbf{11.57}  (0.05) & \textbf{9.02}  (0.03) & 8.05  (0.03) & 6.35  (0.03) \\ 
			\multirow{4}{*}{}&\multirow{4}{*}{}& Pointwise QR & 11.58  (0.05) & 8.98  (0.04) & \textbf{7.90}  (0.03) & 6.31  (0.03) \\ 
			\multirow{4}{*}{}&\multirow{4}{*}{}& Mod CM & 11.69 (0.05) & 9.38 (0.04) & 8.38 (0.04) & 7.01 (0.04) \\ 
			\multirow{4}{*}{}&\multirow{4}{*}{}& PQR & 13.49  (0.05) & 10.95  (0.04) & 8.61  (0.03) & \textbf{5.32}  (0.02) \\ [0.3cm]
			
			\multirow{4}{*}{Mixture}&\multirow{4}{*}{1}& Joint QR &\textbf{13.18}  (0.05) & 10.31  (0.04) & \textbf{8.79}  (0.04) & 6.79  (0.03) \\ 
			\multirow{4}{*}{}&\multirow{4}{*}{}& Pointwise QR & 13.21  (0.05) & \textbf{10.28}  (0.04) & 8.69  (0.04) & 6.73  (0.03) \\ 
			\multirow{4}{*}{}&\multirow{4}{*}{}& Mod CM & 13.26 (0.05) & 10.67 (0.04) & 9.21 (0.04) & 7.47 (0.04) \\ 
			\multirow{4}{*}{}&\multirow{4}{*}{}& PQR & 14.93  (0.06) & 12.05  (0.04) & 9.35  (0.04) & \textbf{5.96}  (0.02) \\ 
			
			\hline\hline
		\end{tabular}		}
		\label{tab:sim1MAE2}
		
	\end{table}
	
	\begin{table}[htbp]
		\caption{Average MAE (standard error in parentheses) of the predicted $\tau$-level quantile for the case of heteroskedasticity with a scalar covariate and a sparsely observed functional covariate.}
		\centering
		\scalebox{0.75}
		{
			\begin{tabular}{ccccccc}
				\hline\hline
				Sample size &  SNR & Method & $\tau=0.05$ & $\tau=0.1$ & $\tau=0.25$ & $\tau=0.5$ \\
				\hline\hline

				\multirow{4}{*}{100}&\multirow{4}{*}{150}& Joint QR & 4.41  (0.03) & 3.99 (0.02) & 3.48 (0.02) & 3.25  (0.02) \\  
				\multirow{4}{*}{}&\multirow{4}{*}{}& Pointwise QR & 5.40 (0.03) & 4.84 (0.03) & 4.26 (0.02) & 4.21  (0.03) \\  
				\multirow{4}{*}{}&\multirow{4}{*}{}& Mod CM & 6.00 (0.03) & 5.71 (0.03) & 5.42 (0.03) & 5.39 (0.03) \\ 
				\multirow{4}{*}{}&\multirow{4}{*}{}& PQR & \textbf{4.38} (0.05) & \textbf{3.43} (0.03) & \textbf{2.52} (0.02) & \textbf{2.14}  (0.02) \\   [0.3cm]
				
				\multirow{4}{*}{100}&\multirow{4}{*}{10}& Joint QR & \textbf{6.93} (0.03) & \textbf{6.40} (0.03) & 5.89 (0.02) & 5.76  (0.02) \\
				\multirow{4}{*}{}&\multirow{4}{*}{}& Pointwise QR & 7.94 (0.04) & 7.20 (0.03) & 6.48 (0.03) & 6.30  (0.03) \\ 
				\multirow{4}{*}{}&\multirow{4}{*}{}& Mod CM & 8.41 (0.04) & 8.22 (0.04) & 8.13 (0.04) & 8.05 (0.04) \\
				\multirow{4}{*}{}&\multirow{4}{*}{}& PQR  & 7.76 (0.06) & 6.57 (0.04) & \textbf{5.47} (0.03) & \textbf{5.08}  (0.02) \\    [0.3cm]
				
				\multirow{4}{*}{100}&\multirow{4}{*}{5}& Joint QR & \textbf{8.44} (0.04) & \textbf{7.72} (0.03) & 7.05 (0.03) & 6.86  (0.03) \\ 
				\multirow{4}{*}{}&\multirow{4}{*}{}& Pointwise QR & 9.44 (0.05) & 8.50 (0.04) & 7.57 (0.03) & 7.26  (0.03) \\ 
				\multirow{4}{*}{}&\multirow{4}{*}{}& Mod CM & 9.65 (0.05) & 9.40 (0.05) & 9.36 (0.05) & 9.26 (0.05) \\
				\multirow{4}{*}{}&\multirow{4}{*}{}& PQR  & 9.51 (0.07) & 8.25 (0.05) & \textbf{6.87} (0.03) & \textbf{6.33} (0.02)  \\    [0.3cm]

				\multirow{4}{*}{100}&\multirow{4}{*}{2}& Joint QR & \textbf{10.68} (0.05) & \textbf{9.63} (0.04) & \textbf{8.60} (0.03) & 8.30  (0.03) \\ 
				\multirow{4}{*}{}&\multirow{4}{*}{}& Pointwise QR &  11.55 (0.06) & 10.29 (0.05) & 9.00 (0.04) & 8.54  (0.03) \\ 
				\multirow{4}{*}{}&\multirow{4}{*}{}& Mod CM & 11.27 (0.05) & 10.80 (0.05) & 10.43 (0.05) & 10.40 (0.06) \\
				\multirow{4}{*}{}&\multirow{4}{*}{}& PQR & 12.07 (0.08) & 10.53 (0.06) & 8.66 (0.04) & \textbf{7.92}  (0.03) \\   [0.3cm]
				
				\multirow{4}{*}{100}&\multirow{4}{*}{1}& Joint QR & \textbf{12.23} (0.06) & \textbf{10.91} (0.05) & \textbf{9.54} (0.04) & 9.11  (0.03) \\ 
				\multirow{4}{*}{}&\multirow{4}{*}{}& Pointwise QR & 12.87 (0.07) & 11.39 (0.05) & 9.85 (0.04) & 9.29  (0.03) \\ 
				\multirow{4}{*}{}&\multirow{4}{*}{}& Mod CM & 12.43 (0.06) & 11.80 (0.06) & 11.19 (0.06) & 11.09 (0.07) \\
				\multirow{4}{*}{}&\multirow{4}{*}{}& PQR &  13.70 (0.09) & 11.91 (0.06) & 9.70 (0.04) & \textbf{8.84}  (0.03) \\  [0.1cm]
				\hline \\  [-0.3cm]

				\multirow{4}{*}{1000}&\multirow{4}{*}{150}& Joint QR & \textbf{2.87} (0.01) & \textbf{2.42} (0.01) & 1.89 (0.01) & 1.65  (0.01) \\ 
				\multirow{4}{*}{}&\multirow{4}{*}{}& Pointwise QR & 3.06 (0.01) & 2.60 (0.01) & 2.07 (0.01) & 1.86  (0.01) \\ 
				\multirow{4}{*}{}&\multirow{4}{*}{}& Mod CM & 3.13 (0.01) & 2.65 (0.01) & 2.10 (0.01) & 1.91 (0.01) \\ 
				\multirow{4}{*}{}&\multirow{4}{*}{}& PQR & 3.10 (0.01) & 2.47 (0.01) & \textbf{1.78} (0.01) & \textbf{1.48} (0.01) \\  [0.3cm]
				
				\multirow{4}{*}{1000}&\multirow{4}{*}{10}& Joint QR & \textbf{6.21} (0.02) & \textbf{5.45} (0.02) & \textbf{4.79} (0.02) & 4.66  (0.02) \\
				\multirow{4}{*}{}&\multirow{4}{*}{}& Pointwise QR & 6.38 (0.02) & 5.59 (0.02) & 4.90 (0.02) & 4.77  (0.02) \\ 
				\multirow{4}{*}{}&\multirow{4}{*}{}& Mod CM & 6.46 (0.02) & 5.70 (0.02) & 5.01 (0.02) & 4.86 (0.02) \\ 
				\multirow{4}{*}{}&\multirow{4}{*}{}& PQR & 6.68 (0.03) & 5.82 (0.02) & 4.92 (0.02) & \textbf{4.60}  (0.02) \\  [0.3cm]
				
				\multirow{4}{*}{1000}&\multirow{4}{*}{5}& Joint QR & \textbf{8.08} (0.03) & \textbf{7.01} (0.02) & \textbf{6.07} (0.02) & 5.85  (0.02) \\
				\multirow{4}{*}{}&\multirow{4}{*}{}& Pointwise QR  & 8.27 (0.03) & 7.15 (0.02) & 6.16 (0.02) & 5.94  (0.02) \\ 
				\multirow{4}{*}{}&\multirow{4}{*}{}& Mod CM & 8.30 (0.03) & 7.26 (0.02) & 6.31 (0.02) & 6.08 (0.02)  \\ 
				\multirow{4}{*}{}&\multirow{4}{*}{}& PQR & 8.64 (0.03) & 7.50 (0.03) & 6.27 (0.02) & \textbf{5.81}  (0.02)  \\  [0.3cm]
				
				\multirow{4}{*}{1000}&\multirow{4}{*}{2}& Joint QR &  \textbf{10.76} (0.03) & \textbf{9.21} (0.03) & \textbf{7.76} (0.02) & 7.38  (0.02) \\ 
				\multirow{4}{*}{}&\multirow{4}{*}{}& Pointwise QR & 10.95 (0.04) & 9.34 (0.03) & 7.85 (0.03) & 7.45  (0.02) \\ 
				\multirow{4}{*}{}&\multirow{4}{*}{}& Mod CM & 10.92 (0.03) & 9.45 (0.03) & 8.01 (0.03) & 7.60 (0.03) \\
				\multirow{4}{*}{}&\multirow{4}{*}{}& PQR & 11.38 (0.04) & 9.79 (0.03) & 8.03 (0.03) & \textbf{7.34}  (0.02) \\ [0.3cm]
				
				\multirow{4}{*}{1000}&\multirow{4}{*}{1}& Joint QR & \textbf{12.54} (0.04) & \textbf{10.65} (0.03) & \textbf{8.83} (0.03) & 8.32  (0.03) \\ 
				\multirow{4}{*}{}&\multirow{4}{*}{}& Pointwise QR & 12.74 (0.04) & 10.81 (0.04) & 8.92 (0.03) & 8.39  (0.03)  \\ 
				\multirow{4}{*}{}&\multirow{4}{*}{}& Mod CM & 12.70 (0.04) & 10.91 (0.04) & 9.10 (0.03) & 8.56 (0.03) \\
				\multirow{4}{*}{}&\multirow{4}{*}{}& PQR & 13.16 (0.05) & 11.26 (0.04) & 9.13 (0.03) & \textbf{8.29}  (0.03) \\ 
				
				\hline\hline
			\end{tabular}
		}
		\label{tab:sim1MAE3}
	\end{table}
	
	\begin{table}[htbp]
		\caption{Average computing time (in seconds) of the three approaches that involve estimating the conditional distribution for the case of having a scalar covariate and a densely observed functional covariate.}
		\centering
		\scalebox{0.8}{
			\begin{tabular}{cccc}
				\hline	\hline
				Distribution & Method & $n = 100$ & $n = 1000$ \\
				\hline	\hline
				\multirow{3}{*}{Normal}& Joint QR & 12 & 133 \\
				\multirow{3}{*}{}& Pointwise QR & 148 & 271 \\
				\multirow{3}{*}{}& Mod CM & 278 & 511 \\ [0.3cm]
				\multirow{3}{*}{Mixture} & Joint QR & 13 & 154 \\
				\multirow{3}{*}{}& Pointwise QR & 151 & 296 \\
				\multirow{3}{*}{}& Mod CM & 327 & 532 \\
				\hline	\hline
			\end{tabular}}
			\label{tab:sim1time}
		\end{table}

		%\FloatBarrier
		\section{Sow data application} \label{sec: sow dat}
		%
		%% % % % % % % % % % % % % % % % % % % % % % % % % % % % % % % % % % % % % % % % % % % % % % % % % % % % % % % % % % % % % % % % % %
		%% % % Describe the experimental study
		%% % % % % % % % % % % % % % % % % % % % % % % % % % % % % % % % % % % % % % % % % % % % % % % % % % % % % % % % % % % % % % % % % %
		%
		Our motivating application is an experimental study carried out at a commercial farm in Oklahoma from July 21, 2013 to August 19, 2013 \citep{rosero2016essential}. The study comprises of 480 lactating sows of different parities (i.e. number of previous pregnancies, which serves as a surrogate for age and body weight) that were observed during their first 21 lactation days; their feed intake was recorded daily as the difference between the feed offer and the feed refusal. In addition the study contains information on the temperature and humidity of the farrowing rooms, each recorded at five minute intervals. The final dataset we used for the analysis consists of 475 sows after five sows with unreliable measurements were removed by the experimenters.
		
		The experiment was conducted to gain better insights into the way that the ambient temperature and humidity of the farrowing room affect the feed intake of lactating sows. Previous studies seem to suggest a reduction in the sow's feed intake due to heat stress:
		above $29^{\circ}$C sows decrease feed intake by 0.5 kg per additional degree in temperature \citep{quiniou1999influence}.
		%%\citet{quiniou1999influence} report that lactating sows severely decrease feed intake by $46\%$ when sows were exposed to $29^{\circ}$C compared to $18^{\circ}$C.
		Studying the effect of heat stress on lactating sows is a very important scientific question because of a couple of reasons. First, the reduction of feed intake of the lactating sows is associated with a decrease in both their bodyweight (BW) and milk production, as well as the weight gain of their litter \citep{johnston1999effect, renaudeau2001effects, renaudeau2001effectsA}. Sows with poor feed intake during lactation continue the subsequent reproductive period with negative energy balance \citep{black1993lactation}, which leads to prevent the onset of a new reproductive cycle.
		%%Second, the heat stress before and after conception also sees a negative effect on the sows reproductive system \citep{st2003economic},
		Second, heat stress reduces farrowing rate (number of sows that deliver a new litter) and number of piglets born \citep{bloemhof2013effect}; the reduction in reproduction due to seasonality is estimated to cost 300 million dollars per year for the swine industry \citep{st2003economic}. Economic losses are estimated to increase \citep{nelson2009climate} because high temperatures are likely to occur more frequently due to global warming \citep{melillo2014climate}.
		
		Our primary goal is to understand the thermal needs of the lactating sows for proper feeding behavior during the lactation time. We are interested in how the interplay between the temperature and humidity of the farrowing room affects the feed intake demeanor of lactating sows of different parities.
		We focus on three specific time points during the lactation period - beginning (lactation day 4), middle (day 11) and end (day 18) - and the analyses are done separately for each time point. We consider two types of responses that are meant to assess the feed intake behavior using the current and the previous lactation days.
		%We focus on three specific times during the lactation period: beginning (lactation day 4), middle (day 11) and end (day 18) and consider two types of responses that are meant to assess the feed intake behavior using the current and the previous lactation days.
		The first one quantifies the absolute change in the feed intake over two consecutive days and the second one quantifies the relative change and takes into account the usual sow's feed intake. We define them formally after introducing some notation.
		
		Let $FI_{ij}$ be the $j$th measurement of the feed intake observed for the $i$th sow and denote by the lactation day $LD_{ij}$ when $FI_{ij}$ is measured; here $j=1, \ldots, n_{i}$. Most sows are observed for every day within the first 21 lactation days and thus have $n_i=21$.
		First define the absolute change in the feed intake between two consecutive days as $\Delta^{(1)}_{i(j+1)} = FI_{i(j+1)} - FI_{ij}$ for $j$ that satisfies $LD_{i(j+1)} - LD_{ij} = 1$. For instance $\Delta^{(1)}_{i(j+1)}=0$ means there was no change in feed intake of sow $i$ between the current day and the previous day, while $\Delta^{(1)}_{i(j+1)} < 0$ means that the feed intake consumed by the $i$th sow in the current day is smaller than the feed intake consumed in the previous day. However, the same amount of change in the feed intake may reflect some stress level for a sow who typically eats a lot and a more serious stress level for a sow that usually has a lower appetite. For this, we define the relative change in the feed intake by $\Delta^{(2)}_{i(j+1)}= (FI_{i(j+1)} - FI_{ij})/$ $\{(LD_{i(j+1)} - LD_{ij}) \cdot TA_{i}\}$, where $TA_{i}$ is the trimmed average of feed intake of $i$th sow calculated as the average feed intake after removing the lowest $20\%$ and highest $20\%$ of the feed intake measurements $\{FI_{i1}, \ldots, FI_{in_i}\}$ taken for the corresponding sow. Here $TA_{i}$ is surrogate for the usual amount of feed intake of the $i$th sow. Trimmed average is used instead of the common average, to remove outliers of very low feed intakes in first few lactating days. For example, consider the situation of two sows:  sow $i$ that \textit{typically} consumes 10lb food per day and  sow $i'$ that consumes 5lb food per day. A reduction of 5lb in the feed intake over two consecutive days corresponds to $\Delta^{(2)}_{i(j+1)} = - 50\%$ for the $i$th sow and  $\Delta^{(2)}_{i'(j+1)} = - 100\%$ for the $i'$th sow. Clearly both sows are stressed (negative value) but the second sow is stressed more, as its absolute relative change is larger; in view of this we refer to the second response as the \emph{stress index}. Due to the construction of the two types of responses the data size varies for lactation days 4 ($j=3$), 11 ($j=10$), and 18 ($j = 17$); for the first response, $\Delta^{(1)}_{i(j+1)}$, we have sample sizes of 233, 350, and 278, whereas for $\Delta^{(2)}_{i(j+1)}$ the sample sizes are 362, 373, and 336 for the respective lactation days.
		%Due to the definition of the two types of responses, the data size varies, so for the first response, $\Delta^{(1)}_{i(j+1)}$, we have sample sizes of 233, 350, and 278 for lactation days 4 ($j=3$), 11 ($j=10$), and 18 ($j=17$), respectively, whereas for $\Delta^{(2)}_{i(j+1)}$ the sample sizes are 362, 373, and 336 respectively.
		
\begin{figure}
			\includegraphics[width=\textwidth]{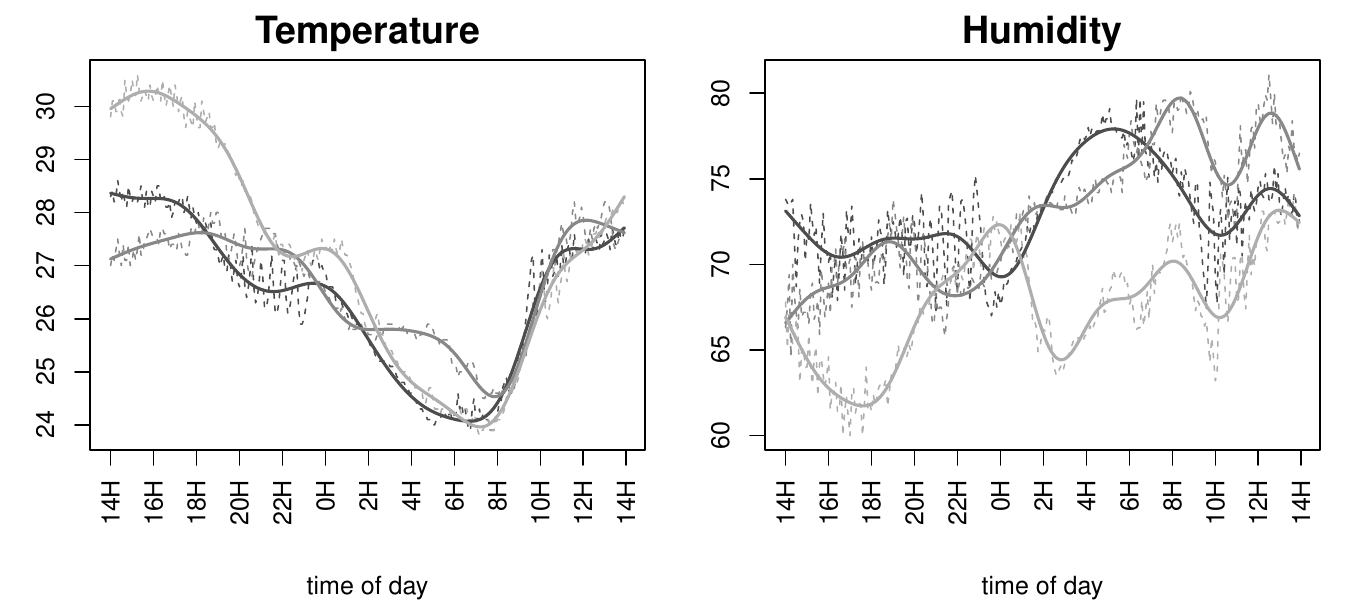} 
			\caption{Temperature ($^{\circ}$C) and humidity ($\%$) observed profiles (dashed) for three randomly selected days and the corresponding smoothed ones (solid); the x-axis begins at 14H (2PM).}
			\label{functional covariates}
		\end{figure}

		In this analysis we center the attention on the effect of the ambient temperature and humidity on the \textit{1st quartile} of the proxy stress measures and gain more understanding of the food consumption of sows that are most susceptible to heat stress.  While the association between the feed intake of lactating sows and the ambient conditions of the farrowing room has been an active research area for some time, accounting for the temperature daily profile has not been considered yet hitherto. Figure \ref{functional covariates}
		displays the temperature and humidity daily profiles recorded at a frequency of 5-minute window intervals for three different days. Preliminary investigation reveals that temperature is negatively correlated with humidity at each time; this phenomenon is caused because the farm uses cool cell panels and fans to control the ambient temperature. Furthermore, it appears that there is a strong pointwise correlation between temperature and humidity. In view of these observations, in our analysis we consider the daily average of humidity. Exploratory analysis of the feed intake behavior of the sows suggest similarities for the sows with parity greater than older sows (ones who are at their third pregnancy or higher); thus we use a parity indicator instead of the actual parity of the sow. The parity indicator $P_i$ is defined as one, if the $i$th sow has parity one and zero otherwise. 
		% REVISIT
		%As emphasized throughout the paper, the existing literature on quantile regression is not suitable to incorporate covariates of different types, as it is the case here.
		%
		
		For the analysis we smooth daily temperature measurements of each sow using univariate smoother with $15$ cubic regression bases and quadratic penalty; REML is used to estimate smoothing parameter. The smoothed temperature curve for sow $i$'s $j$th repeated measure is denoted by $T_{ij}(t)$, $t\in[0,24)$, and the corresponding daily average humidity is denoted by $AH_{ij}$. Both temperature and average humidity are centered before being used in the analysis.

		For convenience we denote the response with $\Delta_{ij}$ by removing the superscript. In this application for fixed $j$, $\Delta_{ij}$ corresponds to $Y_{i}$ in Section \ref{sec: method}, $P_i$ and $AH_{ij}$ correspond to scalar covariates $X_{i2}$, and $T_{ij}(t)$ and $AH_{ij} \cdot T_{ij}(t)$ to functional covariates $X_{i3}(\cdot)$.  
		%In the following we detail estimation procedure. Since the procedure is identical for both responses here we remove superscript in notation and use $\Delta_{ij}$ as our response. 
		We first estimate the conditional distribution of $\Delta_{ij}$ given temperature $T_{ij}(t)$, average humidity $AH_{ij}$, parity $P_{i}$, and interaction $AH_{ij} \cdot T_{ij}(t)$. Specifically for each of lactation days of interest ($j = 3, 10$ and $17$) we create a set of $100$ equi-spaced grid of points between the fifth smallest and fifth largest values of $\Delta_{ij}$'s and denote the grids with $\mathcal{D} = \{d_{\ell} : \ell = 1, \ldots, 100\}$. Then we create artificial binary responses, $\{\mathbbm{1}\left(\Delta_{ij} \leq d_{\ell}\right): \ell = 1, \ldots, 100\}$, and fit the following model for $F_{ij}(d_{\ell}) = E\left[\mathbbm{1}\left(\Delta_{ij} \leq d_{\ell}\right) \big| T_{ij}(t), AH_{ij}, P_{i}\right]$:
		\begin{align*}
			E\left[\mathbbm{1}\left(\Delta_{ij}  \leq  d_{\ell}\right) \big| T_{ij}(t), AH_{ij}, P_{i}\right]  \nonumber  = g^{-1}  \Big \{ \beta_0(d_{\ell}) + \beta_1(d_{\ell}) P_i   + \beta_2(d_{\ell}) AH_{ij} \\
			+ \int\beta_3(d_{\ell},t)T_{ij}(t)dt + AH_{ij} \int\beta_4(d_{\ell},t)T_{ij}(t)dt  \Big \}, \nonumber
		\end{align*}
		where $\beta_0(\cdot)$ is a smooth intercept, $\beta_1(\cdot)$ quantifies the smooth effect of young sows, $\beta_2(\cdot)$ describes the effect of the humidity,  and  $\beta_3(\cdot, t)$ and $\beta_4(\cdot,t)$ quantify the effect of the temperature at time $t$ as well as the interaction between the temperature at time $t$ and average humidity. We model $\beta_0(\cdot)$ using 20 univariate basis functions, $\beta_1(\cdot)$ and $\beta_2(\cdot)$ using five univariate basis functions, $\beta_3(\cdot,\cdot)$ and $\beta_4(\cdot,\cdot)$ using tensor product of two univariate bases functions (total of 25 functions). Throughout the analysis, cubic B-spline bases are used and REML is used for estimating smoothing parameters. The estimated conditional distribution, denoted by $\widehat{F}_{ij}(d)$, is monotonized by fitting isotonic regression to $\{(d_{\ell}, \widehat{F}_{ij}(d_{\ell})):\ell = 10, \ldots, 90\}$ ; ten smallest and ten largest $d_{\ell}$ and the corresponding values of $\widehat{F}_{j}(d_\ell)$ are removed to avoid boundary effects. By abuse of notation, $\widehat{F}_{ij}(d)$ denotes the resulting monotonized estimated distribution. Finally, we obtain estimated first quartiles, i.e. quantiles at $\tau=0.25$ level, by inverting $\widehat{F}_{ij}(d)$, namely $\widehat{Q}\left(\tau = 0.25 \; | \; T_{ij}(t), AH_{ij}, P_{i} \right) = \text{inf}\{d : \widehat{F}_{ij}(d) \geq 0.25\}$.
		
		To understand the relationship between the lactating sows feed intake and the thermal condition of the farrowing room, we systematically compare and study the predicted quantiles of two responses at combinations of different values of temperature, humidity, and parity. For each of three lactation days ($j = 3, 10, 17$) we consider three values of average humidity (first quartile, median, and third quartile) and two levels of parity ($0$ for older sows and $1$ for younger sows). Based on the experimenters' interest, for the functional covariate $T_{ij}(\cdot)$ we consider seven smooth temperature curves given in Figure 2. Each of these curves are obtained by first calculating pointwise quantiles of temperature at five-minute intervals for a specific level and then smoothing it; we considered quantiles levels $\eta = 0.2, 0.3, \ldots,$ and $0.8$. In short, for each of three lactation days we obtain the first quartile of two responses for 42 different combinations ($3$ humidity values $\times$ 2 parity levels $\times$ 7 temperature curves) using the proposed method. To avoid extrapolation we ascertain that (i) there are reasonably many observed measurements at each of the combinations and (ii) bottom 25\% of the responses are not dominantly from one of the parity group; see distribution of each response by the parity in Figure \ref{response_by_parity}.
		
		%To study and interpret the effect of each covariate we predict quantiles at combinations of different values of covariates and investigate the predicted values against the covariates. For each of three lactation days ($ j = 3, 10, 17$) we consider three values of average humidity (first quartile, median, and third quartile) and the two levels of parity ($0$ for older sows and $1$ for younger sows). For the functional covariate $T_{ij}(\cdot)$ we create seven smooth temperature curves given in Figure \ref{temperature curves} by first obtaining pointwise quantiles of temperature and then smoothing them for each of quantiles levels $\eta = 0.2, 0.3, \ldots,$ and $0.8$. In summary for each of three lactation days we make prediction of distribution and obtain the first quartile of two responses for $42$ ($3 \times 2 \times 7$) different combinations of covariates values. To avoid extrapolation we ascertain that there are reasonably many observed measurements at each of the combinations and bottom $25\%$ of the responses are not dominantly from one of the parity group; see distribution of each response by the parity in Figure \ref{response_by_parity}.
		
		\begin{figure}
			\centering
			\includegraphics[width = 0.9\textwidth]{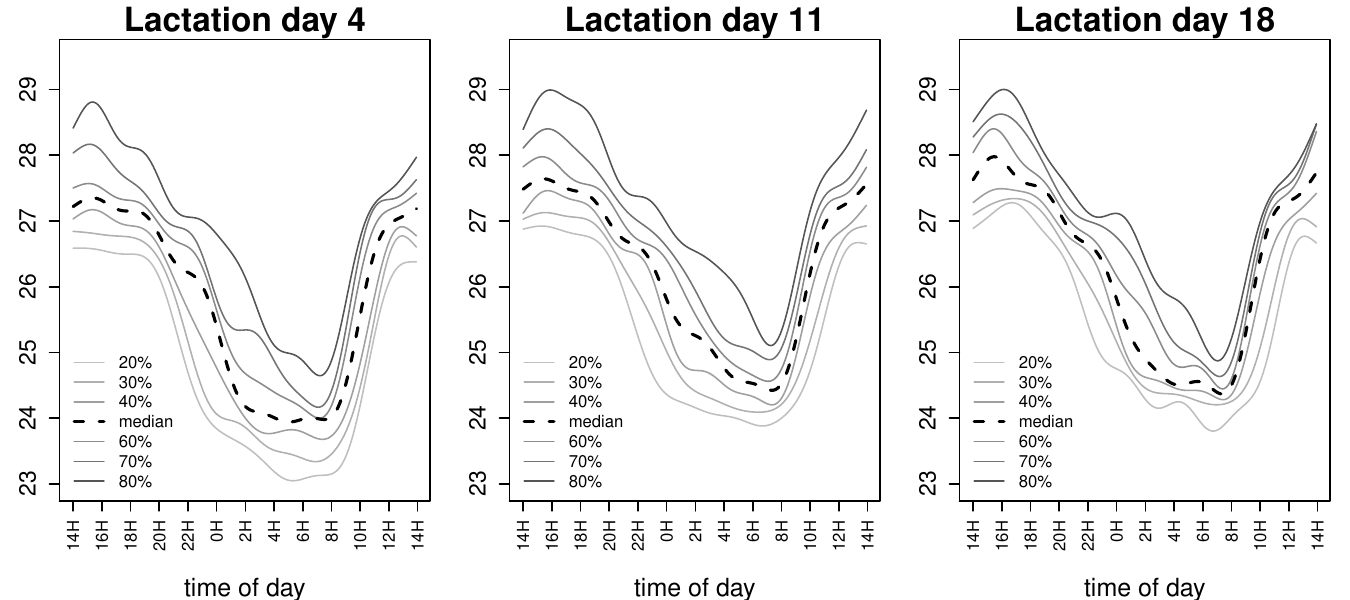}
			\caption{Temperature curves with which prediction of quantiles is made. Dashed black line is pointwise average of temperature curves and solid lines are pointwise quartiles; all curves are smoothed.}
			\label{temperature curves}
		\end{figure}
		\begin{figure}
			\centering
			\includegraphics[width = \textwidth]{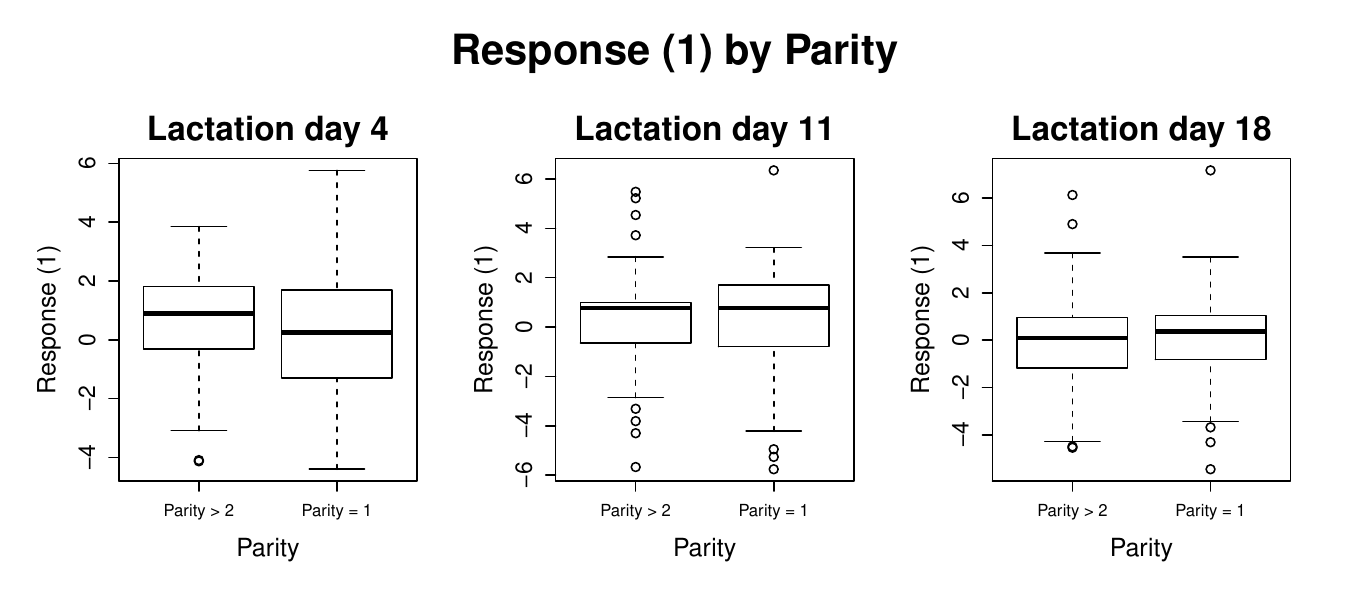}
			\includegraphics[width = \textwidth]{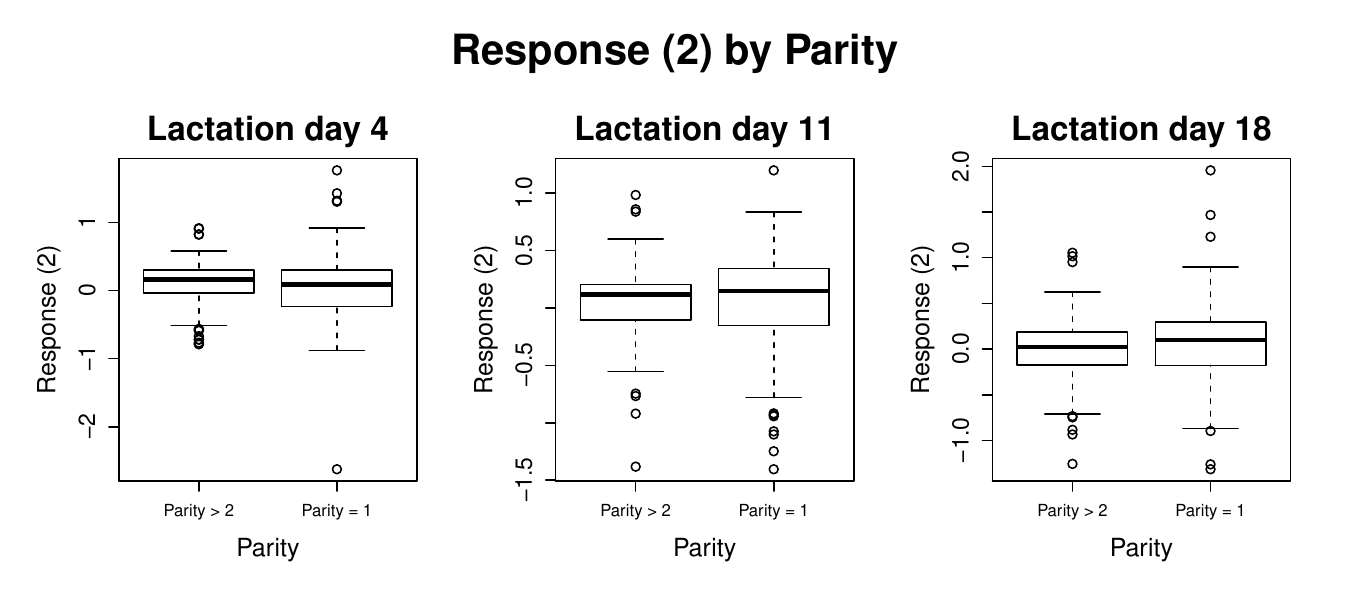}
			\caption{Top panels: back-to-back boxplots of the absolute change in feed intake at a specific day by parity; Bottom panels: back-to-back boxplots of the relative change in feed intake at a specific day by parity.}
			\label{response_by_parity}
		\end{figure}
		
\begin{figure}
			\centering
			\includegraphics[width = 0.8\textwidth]{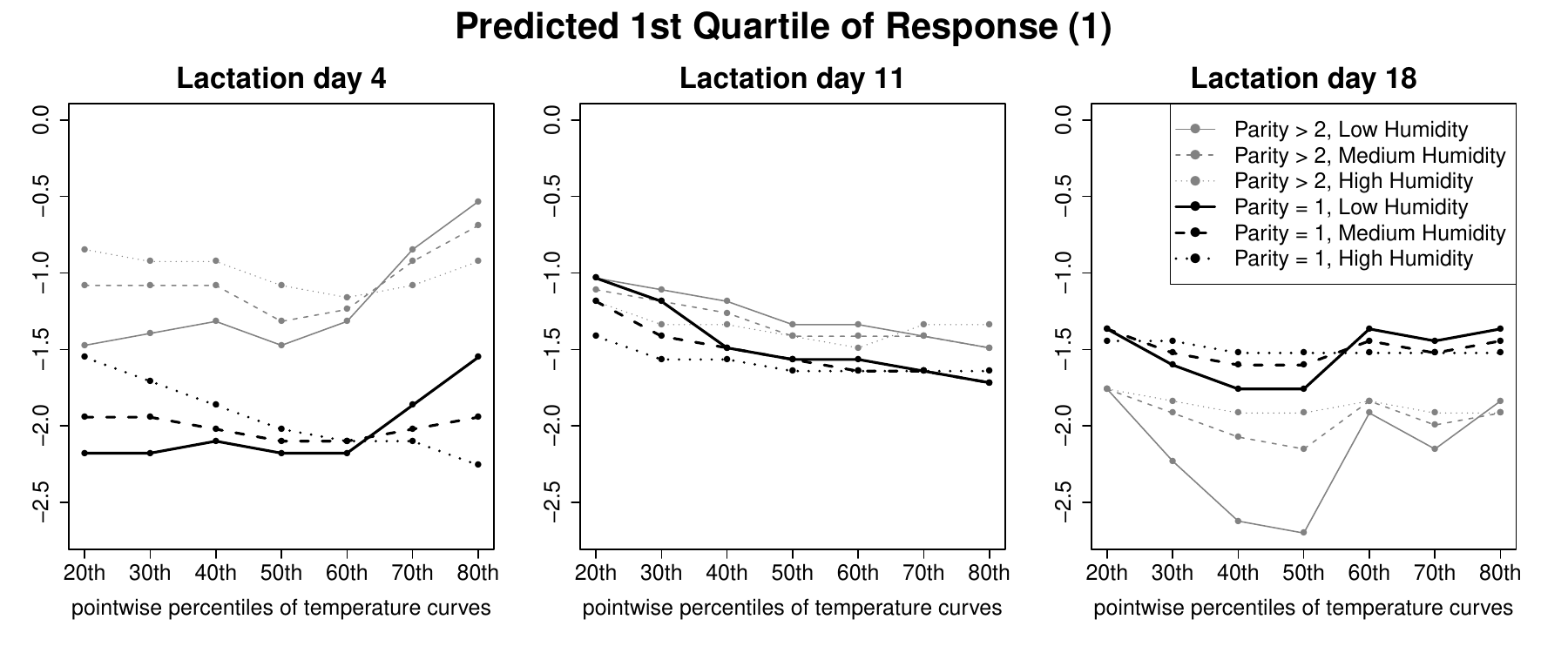}
			\includegraphics[width = 0.8\textwidth]{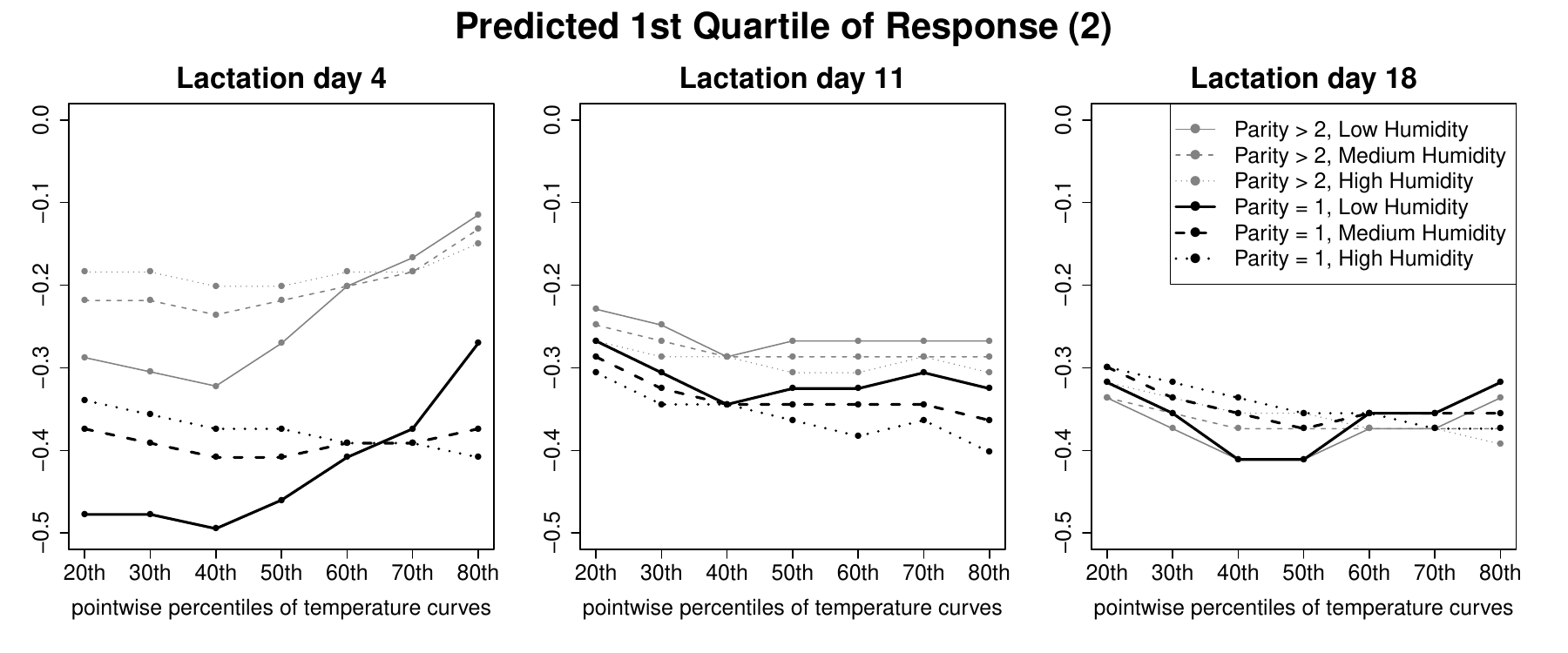}
			\caption{Displayed are the predicted quantiles of $\Delta^{(1)}_{i(j+1)}$ and $\Delta^{(2)}_{i(j+1)}$ for different parities, average humidity, and temperature levels. In each of all six panels, black thick lines correspond to the young sows ($P_i = 1$) and grey thin lines correspond to the old sows ($P_{i} = 0$). Line types indicate different average humidity levels; solid, dashed, and dotted correspond to low, medium, and high average humidity levels (given by the first quartile, median, and the third quartiles of $AH_{ij}$), respectively. The seven grids in $x$-axis of each panel correspond to the $7$ temperature curves given in the respective panel of Figure \ref{temperature curves}.}
			\label{predicted quantile}
		\end{figure}

		The resulting predicted quantiles are shown in Figure \ref{predicted quantile}. Here we focus our discussion on predicted quantile of $\Delta^{(2)}_{i(j+1)}$ at quantile level $\tau = 0.25$ for lactation day 4 ($j=3$) - the first plot of the second row in Figure \ref{predicted quantile}. The results suggest that the feed intake of older sows (parity $P_i = 0$; grey lines) are less affected by high temperatures than younger sows (black lines); this finding is in agreement with
		\citet{bloemhof2013effect}. We also observe that the effects of humidity and temperature on feed intake change are strongly intertwined. For illustration, we focus on lactation day 4 ($j = 3$) again for younger sows (black lines). For medium humidity (dashed lines) their feed intake stays pretty constant as temperature increases, while for low and high humidity levels (solid and dotted lines, respectively) it changes with an opposite direction. Specifically when temperature increases the predicted first quartile of $\Delta^{(2)}_{i(j+1)}$ increases for low humidity (solid line) whereas it decreases for high humidity (dotted line). Our results imply that high humidity (dotted line) is related to a negative impact of high temperature on feed intake while low humidity (solid line) alleviates it; and this finding is consistent with a previous study \citep{bergsma2012exploring}. The analysis result suggests to keep low humidity levels in order to maintain healthy feed intake behavior, when ambient temperature is above $60$th percentile; high humidity levels are desirable for cooler ambient temperature. 
		
		Interpretation of the other results is similar.
		While the effects of covariates on feed intake are less apparent toward the end of lactation period, we still observe similar pattern across all three lactation days. For the 11th day ($j = 10$), the 25th quantile of the feed intake is predicted to decrease when the temperature stays below the $40$th percentile, regardless of humidity level and sows age. However, it starts increasing with low humidity while it continues decreasing with high humidity when the temperature rises above the $40$th percentile.
		%For lactation day 11 ($j=10$) we observe that when the temperature rises above the $40$th percentile, the predicted first quartile starts increasing with low humidity while it continues decreasing with high humidity. 
		Similarly, for the 18th day ($j=17$) when the temperature rises above the $60$th percentile, the predicted first quartile increases with low humidity while it decreases with high humidity. 
		The effect of temperature on feed intake seems less obvious for lactation days 11 and 18 than for day 4; while the effect may be due to lactation day, it may also be a result of other factors, such as more fluctuation and variability in temperature curves on day 4 than on other two days (see Figure \ref{temperature curves}). Overall we conclude that high humidity and temperature affect the sows feed intake behavior negatively and young sows (parity one) are more sensitive to heat stress than older sows (higher parity), especially at the beginning of lactation period.

		%	\vspace{-2em}
		%\FloatBarrier
		
		\section{Discussion}\label{sec: dis}
		
		%We proposed a novel framework for a comprehensive study of covariates of mixed types on the conditional distribution of the response. The modeling approach is flexible and easy to implement and leads to good computational efficiency. Extensive simulation study showed excellent prediction performance in terms of predicting quantiles of various levels.
		
		The proposed modeling framework opens up a couple of future research directions. A first research avenue is to develop significance tests of null covariate effect. Testing for the null effect of a covariate on the conditional distribution of the response is equivalent to testing that the corresponding regression coefficient function is equal to zero in the associated function-on-function mean regression model. Such significance tests have been studied when the functional response is continuous \citep{shen2004f, zhang2007statistical}; however their study for binary-valued functional responses is an open problem in functional data literature, and only recently has been considered in \cite{chen2018}.
		%One possible alternative is to construct confidence bands for the corresponding coefficient function, say using bootstrap, and examine whether the confidence band includes zero for its entire domain.
		Another research avenue is to do variable selection in the setting where there are many scalar covariates and functional covariates. Many current applications collect data with increasing number of mixed covariates and selecting the ones that have an effect on the conditional distribution of the response is very important. This problem is an active research area in functional mean regression where the response is normal \citep{gertheiss2013variable, chen2015variable}. The proposed modeling framework has the potential to facilitate studying such problem.

		%\begin{supplement}
		%\sname{}
		%\stitle{}
		%\slink[url]{/SuppMat.pdf}
		%\sdescription{Section \ref{sec: sim-add} provides additional simulation settings and results for the cases of having either a single scalar covariate or a single functional covariate. Section \ref{sec: bike sharing} presents additional data analysis done using the proposed method on the bike sharing dataset \citep{BSD2013, Lichman:2013}. Lastly the \texttt{R} function implementing the proposed method is available at \url{ http://www4.ncsu.edu/~spark13/software/QRFD_Rcode.zip/ }. }
		%\end{supplement}
		
		\section*{Data Availability}
		The data used to support the findings of this study are available from the corresponding author upon request.
		
		\section*{Conflicts of Interest}
		The authors declare that they have no conflicts of interests.
		
		\section*{Funding Statement}
		Staicu's research was supported  by National Science Foundation DMS 0454942 and DMS 1454942, and National Institutes of Health grants R01 NS085211, R01 MH086633, 5P01 CA142538-09. 
		
		\section*{Acknowledgement}
		%Staicu's research was supported by National Science Foundation DMS 0454942 and DMS 1454942, and National Institutes of Health grants R01 NS085211, R01 MH086633, 5P01 CA142538-09. 
		The data used originated from work supported in part by the North Carolina Agricultural Foundation, Raleigh, NC. The authors acknowledge Zhen Han for preparing simulations. 
		
		%Staicu's research was supported by National Science Foundation grant number DMS 1454942 and National Institute of Health grant number 5P01 CA142538-09. 
		
		\section*{Supplementary materials}
		Section \ref{sec: sim-add} provides additional simulation settings and results for the cases of having either a single scalar covariate or a single functional covariate. Additional results for the case of having a scalar covariate and a densely observed functional covariate are also included.
		Section \ref{sec: bike sharing} presents additional data analysis using the proposed method on the bike sharing dataset \citep{BSD2013, Lichman:2013}.

\bibliography{ref}	

\begin{thebibliography}{}

\bibitem[Barlow, 1972]{barlow1972statistical}
Barlow, R.~E. (1972).
\newblock Statistical inference under order restrictions; the theory and
  application of isotonic regression.
\newblock Technical report.

\bibitem[Bergsma and Hermesch, 2012]{bergsma2012exploring}
Bergsma, R. and Hermesch, S. (2012).
\newblock Exploring breeding opportunities for reduced thermal sensitivity of
  feed intake in the lactating sow.
\newblock {\em Journal of animal science}, 90(1):85--98.

\bibitem[Black et~al., 1993]{black1993lactation}
Black, J., Mullan, B., Lorschy, M., and Giles, L. (1993).
\newblock Lactation in the sow during heat stress.
\newblock {\em Livestock production science}, 35(1):153--170.

\bibitem[Bloemhof et~al., 2013]{bloemhof2013effect}
Bloemhof, S., Mathur, P., Knol, E., and Van~der Waaij, E. (2013).
\newblock Effect of daily environmental temperature on farrowing rate and total
  born in dam line sows.
\newblock {\em Journal of animal science}, 91(6):2667--2679.

\bibitem[Bondell et~al., 2010]{BondellReichWang2010}
Bondell, H., Reich, B., and Wang, H. (2010).
\newblock Noncrossing quantile regression curve estimation.
\newblock {\em Biometrika}, 97(4):825--838.

\bibitem[Cardot et~al., 2005]{cardot2005quantile}
Cardot, H., Crambes, C., and Sarda, P. (2005).
\newblock Quantile regression when the covariates are functions.
\newblock {\em Nonparametric Statistics}, 17(7):841--856.

\bibitem[Cardot et~al., 1999]{CardotFerratySarda1999}
Cardot, H., Ferraty, F., and Sarda, P. (1999).
\newblock Functional linear model.
\newblock {\em Statistics \& Probability Letters}, 45(1):11--22.

\bibitem[Chen and M{\"u}ller, 2012]{chen2012conditional}
Chen, K. and M{\"u}ller, H.-G. (2012).
\newblock Conditional quantile analysis when covariates are functions, with
  application to growth data.
\newblock {\em Journal of the Royal Statistical Society: Series B (Statistical
  Methodology)}, 74(1):67--89.

\bibitem[Chen et~al., 2018]{chen2018}
Chen, S., Xiao, L., and Staicu, A.-M. (2018).
\newblock Testing for generalized scalar-on-function linear models.
\newblock Submitted.

\bibitem[Chen et~al., 2016]{chen2015variable}
Chen, Y., Goldsmith, J., and Ogden, R.~T. (2016).
\newblock Variable selection in function-on-scalar regression.
\newblock {\em Stat}, 5(1):88--101.

\bibitem[Chernozhukov et~al., 2009]{chernozhukov2009improving}
Chernozhukov, V., Fernandez-Val, I., and Galichon, A. (2009).
\newblock Improving point and interval estimators of monotone functions by
  rearrangement.
\newblock {\em Biometrika}, 96(3):559--575.

\bibitem[Eilers and Marx, 1996]{eilers1996flexible}
Eilers, P.~H. and Marx, B.~D. (1996).
\newblock Flexible smoothing with b-splines and penalties.
\newblock {\em Statistical science}, pages 89--102.

\bibitem[Fanaee-T and Gama, 2014]{BSD2013}
Fanaee-T, H. and Gama, J. (2014).
\newblock Event labeling combining ensemble detectors and background knowledge.
\newblock {\em Progress in Artificial Intelligence}, 2(2-3):113--127.

\bibitem[Ferraty et~al., 2005]{ferraty2005conditional}
Ferraty, F., Rabhi, A., and Vieu, P. (2005).
\newblock Conditional quantiles for dependent functional data with application
  to the climatic ``el ni{\~n}o" phenomenon.
\newblock {\em Sankhy{\=a}: The Indian Journal of Statistics}, pages 378--398.

\bibitem[Ferraty and Vieu, 2006]{ferraty2006nonparametric}
Ferraty, F. and Vieu, P. (2006).
\newblock {\em Nonparametric Functional Data Analysis}.
\newblock Springer, New York.

\bibitem[Ferraty and Vieu, 2009]{ferraty2009additive}
Ferraty, F. and Vieu, P. (2009).
\newblock Additive prediction and boosting for functional data.
\newblock {\em Computational Statistics \& Data Analysis}, 53(4):1400--1413.

\bibitem[Gertheiss et~al., 2013]{gertheiss2013variable}
Gertheiss, J., Maity, A., and Staicu, A.-M. (2013).
\newblock Variable selection in generalized functional linear models.
\newblock {\em Stat}, 2(1):86--101.

\bibitem[Goldsmith et~al., 2011]{goldsmith2012penalized}
Goldsmith, J., Bobb, J., Crainiceanu, C.~M., Caffo, B., and Reich, D. (2011).
\newblock Penalized functional regression.
\newblock {\em Journal of Computational and Graphical Statistics},
  20(4):830--851.

\bibitem[Huang et~al., 2015]{Rrefund}
Huang, L., Scheipl, F., Goldsmith, J., Gellar, J., Harezlak, J., Mclean, M.,
  Swihart, B., Xiao, L., Crainiceanu, C., Reiss, P., Chen, Y., Greven, S., Huo,
  L., Kundu, M., and Wrobel, J. (2015).
\newblock {\em {R package {\it refund}: Methodology for regression with
  functional data (version 0.1-13)}}.
\newblock URL: \url{https://cran.r-project.org/web/packages/refund/index.html}.

\bibitem[Ivanescu et~al., 2015]{ivanescu2014penalized}
Ivanescu, A.~E., Staicu, A.-M., Scheipl, F., and Greven, S. (2015).
\newblock Penalized function-on-function regression.
\newblock {\em Computational Statistics}, 30(2):539--568.

\bibitem[James, 2002]{james2002generalized}
James, G.~M. (2002).
\newblock Generalized linear models with functional predictors.
\newblock {\em Journal of the Royal Statistical Society: Series B (Statistical
  Methodology)}, 64(3):411--432.

\bibitem[Johnston et~al., 1999]{johnston1999effect}
Johnston, L., Ellis, M., Libal, G., Mayrose, V., and Weldon, W. (1999).
\newblock Effect of room temperature and dietary amino acid concentration on
  performance of lactating sows. ncr-89 committee on swine management.
\newblock {\em Journal of animal science}, 77(7):1638--1644.

\bibitem[Kato, 2012]{kato2012estimation}
Kato, K. (2012).
\newblock Estimation in functional linear quantile regression.
\newblock {\em The Annals of Statistics}, 40(6):3108--3136.

\bibitem[Kim et~al., 2018]{kim2018additive}
Kim, J.~S., Staicu, A.-M., Maity, A., Carroll, R.~J., and Ruppert, D. (2018).
\newblock Additive function-on-function regression.
\newblock {\em Journal of Computational and Graphical Statistics},
  27(1):234--244.

\bibitem[Koenker, 2005]{koenker2005quantile}
Koenker, R. (2005).
\newblock {\em Quantile Regression}.
\newblock Number~38. Cambridge university press.

\bibitem[Koenker and Bassett~Jr, 1978]{koenker1978regression}
Koenker, R. and Bassett~Jr, G. (1978).
\newblock Regression quantiles.
\newblock {\em Econometrica: journal of the Econometric Society}, pages 33--50.

\bibitem[Lichman, 2013]{Lichman:2013}
Lichman, M. (2013).
\newblock {UCI} machine learning repository.

\bibitem[Lu et~al., 2014]{lu2014functional}
Lu, Y., Du, J., and Sun, Z. (2014).
\newblock Functional partially linear quantile regression model.
\newblock {\em Metrika}, 77(2):317--332.

\bibitem[McLean et~al., 2014]{McLean2014}
McLean, M.~W., Hooker, G., Staicu, A.-M., Scheipl, F., and Ruppert, D. (2014).
\newblock Functional generalized additive models.
\newblock {\em Journal of Computational and Graphical Statistics},
  23(1):249--269.

\bibitem[Melillo, 2014]{melillo2014climate}
Melillo, J.~M. (2014).
\newblock Climate change impacts in the united states, highlights: Us national
  climate assessment.

\bibitem[Nelson et~al., 2009]{nelson2009climate}
Nelson, G.~C., Rosegrant, M.~W., Koo, J., Robertson, R., Sulser, T., Zhu, T.,
  Ringler, C., Msangi, S., Palazzo, A., Batka, M., et~al. (2009).
\newblock {\em Climate Change: Impact on Agriculture and Costs of Adaptation},
  volume~21.
\newblock Intl Food Policy Res Inst.

\bibitem[Ng and Maechler, 2007]{Ng2007cobs}
Ng, P. and Maechler, M. (2007).
\newblock A fast and efficient implementation of qualitatively constrained
  quantile smoothing splines.
\newblock {\em Statistical Modelling}, 7(4):315--328.

\bibitem[Quiniou and Noblet, 1999]{quiniou1999influence}
Quiniou, N. and Noblet, J. (1999).
\newblock Influence of high ambient temperatures on performance of multiparous
  lactating sows.
\newblock {\em Journal of animal science}, 77(8):2124--2134.

\bibitem[Ramsay and Silverman, 2005]{ramsay2006functional}
Ramsay, J. and Silverman, B. (2005).
\newblock {\em Functional Data Analysis}.
\newblock Springer, New York.

\bibitem[Ramsay and Silverman, 2002]{ramsay2002applied}
Ramsay, J. and Silverman, B.~W. (2002).
\newblock {\em Applied Functional Data Analysis: Methods and Case Studies}.
\newblock Springer, New York.

\bibitem[Renaudeau and Noblet, 2001]{renaudeau2001effects}
Renaudeau, D. and Noblet, J. (2001).
\newblock Effects of exposure to high ambient temperature and dietary protein
  level on sow milk production and performance of piglets.
\newblock {\em Journal of animal science}, 79(6):1540--1548.

\bibitem[Renaudeau et~al., 2001]{renaudeau2001effectsA}
Renaudeau, D., Quiniou, N., and Noblet, J. (2001).
\newblock Effects of exposure to high ambient temperature and dietary protein
  level on performance of multiparous lactating sows.
\newblock {\em Journal of Animal Science}, 79(5):1240--1249.

\bibitem[Rosero et~al., 2016]{rosero2016essential}
Rosero, D.~S., Boyd, R.~D., McCulley, M., Odle, J., and van Heugten, E. (2016).
\newblock Essential fatty acid supplementation during lactation is required to
  maximize the subsequent reproductive performance of the modern sow.
\newblock {\em Animal Reproduction Science}, 168:151--163.

\bibitem[Ruppert, 2002]{ruppert2002selecting}
Ruppert, D. (2002).
\newblock Selecting the number of knots for penalized splines.
\newblock {\em Journal of computational and graphical statistics},
  11(4):735--757.

\bibitem[Scheipl et~al., 2016]{scheipl2015generalized}
Scheipl, F., Gertheiss, J., Greven, S., et~al. (2016).
\newblock Generalized functional additive mixed models.
\newblock {\em Electronic Journal of Statistics}, 10(1):1455--1492.

\bibitem[Shen and Faraway, 2004]{shen2004f}
Shen, Q. and Faraway, J. (2004).
\newblock An f test for linear models with functional responses.
\newblock {\em Statistica Sinica}, pages 1239--1257.

\bibitem[St-Pierre et~al., 2003]{st2003economic}
St-Pierre, N., Cobanov, B., and Schnitkey, G. (2003).
\newblock Economic losses from heat stress by us livestock industries.
\newblock {\em Journal of dairy science}, 86:E52--E77.

\bibitem[Tang and Cheng, 2014]{tang2014partial}
Tang, Q. and Cheng, L. (2014).
\newblock Partial functional linear quantile regression.
\newblock {\em Science China Mathematics}, 57(12):2589--2608.

\bibitem[Wood, 2006]{wood2006}
Wood, S. (2006).
\newblock {\em Generalized Additive Models: An Introduction with R}.
\newblock Chapman and Hall/CRC.

\bibitem[Wood, 2003]{wood2003thin}
Wood, S.~N. (2003).
\newblock Thin plate regression splines.
\newblock {\em Journal of the Royal Statistical Society: Series B (Statistical
  Methodology)}, 65(1):95--114.

\bibitem[Xiao et~al., 2018a]{xiao2016face}
Xiao, L., Li, C., Checkley, W., and Crainiceanu, C. (2018a).
\newblock Fast covariance estimation for sparse functional data.
\newblock {\em Statistics and computing}, 28(3):511--522.

\bibitem[Xiao et~al., 2018b]{Xiao:16c}
Xiao, L., Li, C., Checkley, W., and Crainiceanu, C. (2018b).
\newblock {\em R package {\it face}: Fast covariance estimation for sparse
  functional data (version 0.1-4)}.
\newblock URL:\url{http://cran.r-project.org/web/packages/face/index.html}.

\bibitem[Xiao et~al., 2016]{xiao2016fast}
Xiao, L., Zipunnikov, V., Ruppert, D., and Crainiceanu, C. (2016).
\newblock Fast covariance estimation for high-dimensional functional data.
\newblock {\em Statistics and Computing}, 26(1-2):409--421.

\bibitem[Yao et~al., 2005]{yao2005functional}
Yao, F., M{\"u}ller, H.-G., and Wang, J.-L. (2005).
\newblock Functional data analysis for sparse longitudinal data.
\newblock {\em Journal of the American Statistical Association},
  100(470):577--590.

\bibitem[Yu et~al., 2016]{yu2015partial}
Yu, D., Kong, L., and Mizera, I. (2016).
\newblock Partial functional linear quantile regression for neuroimaging data
  analysis.
\newblock {\em Neurocomputing}, 195:74--87.

\bibitem[Zhang and Chen, 2007]{zhang2007statistical}
Zhang, J.-T. and Chen, J. (2007).
\newblock Statistical inferences for functional data.
\newblock {\em The Annals of Statistics}, 35(3):1052--1079.

\end{thebibliography}
\bibliographystyle{apalike}

\end{document}